\documentclass[12pt,english,nofootinbib]{revtex4-2}
\usepackage{lmodern}

\usepackage[T1]{fontenc}
\usepackage[latin9]{inputenc}
\usepackage{babel}
\usepackage{array}
\usepackage{multirow}
\usepackage{dsfont}
\usepackage{amsmath}
\usepackage{amsthm}
\usepackage{amssymb}
\usepackage{graphicx}
\usepackage{geometry}
\usepackage{wasysym}
\usepackage[pdfusetitle,
 bookmarks=true,bookmarksnumbered=false,bookmarksopen=false,
 breaklinks=false,pdfborder={0 0 0},pdfborderstyle={},backref=false,colorlinks=false]
{hyperref}

\makeatletter

\usepackage{diagbox}

\makeatother

\begin{document}
\title{Radial adiabatic perturbations of stellar compact objects}
\author{Paulo Luz}
\email{paulo.luz@tecnico.ulisboa.pt}

\affiliation{Centro de Astrof\'isica e Gravita\c{c}\~ao - CENTRA, Departamento
de F\'isica, Instituto Superior T\'ecnico - IST, Universidade de
Lisboa - UL, Av. Rovisco Pais 1, 1049-001 Lisboa, Portugal}
\affiliation{Escola Superior N\'{a}utica Infante D.	Henrique, Pa\c{c}o de Arcos, Portugal.}

\author{Sante Carloni}
\email{sante.carloni@unige.it}

\affiliation{Institute of Theoretical Physics, Faculty of Mathematics and Physics,
Charles University, Prague, V Hole\v{s}ovi\v{c}k\'ach 2, 180 00
Prague 8, Czech Republic,}
\affiliation{DIME, Universit\`a di Genova, Via all'Opera Pia 15, 16145 Genova,
Italy,}
\affiliation{INFN Sezione di Genova, Via Dodecaneso 33, 16146 Genova, Italy}
\begin{abstract}
We present a covariant and gauge-invariant formulation of the theory of radial adiabatic linear perturbations of self-gravitating, non-dissipative
imperfect fluids within the theory of general relativity.
By codifying the thermodynamical properties of the source into an equation of state and an ansatz on anisotropic pressure that involves  both matter and
kinematic variables, we obtain a set of equations that is directly
applicable to a wide variety of thermodynamic theories for matter
fields. As examples, we evaluate and compare the predictions of the Eckart theory, the Bemfica-Disconzi-Noronha-Kovtun theory, and the Truncated Israel-Stewart
theory on the properties and evolution of radial adiabatic perturbations
of stellar compact objects modeled by classical equilibrium solutions.
Introducing a new solution of the Einstein field equations, and
imposing causality, we propose an upper bound for the maximum compactness
of dynamically stable stars with non-trivial radial and tangential pressures.
\end{abstract}
\maketitle

\section{Introduction}

The structure and evolution of compact stellar objects are inextricably connected with the nature of the fluid source. This is particularly true when considering the effects of anisotropic stresses, which, in this specific context, characterize pressures that do not act toward the exterior of the stellar object. Indeed, studies of the innermost structure of matter reveal that anisotropic stresses are necessary to accurately describe matter's behavior. In addition, isotropic matter together with strong electromagnetic fields can also be globally represented as an anisotropic fluid (see, e.g., \cite{Herrera:1997plx, Cadogan:2024mcl,Cadogan:2024ohj,Cadogan:2024ohj,Cadogan:2024ywc} and references therein).

With the advent of multimessenger astrophysics, gravitational waves have become a crucial probe of the universe. This makes it increasingly important to understand the detailed physics of potential gravitational wave sources, such as neutron stars. In particular, accurately modeling shearing stresses in compact stellar objects and their impact on perturbations is essential.

In recent years, there have been several studies on the solutions of the Tolman-Oppenheimer-Volkoff  (TOV) equations sourced by anisotropic fluids, considered under several different perspectives~\cite{Herrera_Barreto_2013,Herrera:2007kz,Chaisi:2006sc,Herrera:2015vca,Harko:2002pxr,Hoffberg:1970vqj,Rago:1991qe,Richardson:1972xn,Viaggiu:2008yf,Thirukkanesh:2018hfy,Estrada:2018zbh,Singh:2016mqs,Maurya:2016oml,Maurya:2017gth,Ivanov:2017kyr,Jasim:2016cmk,Maurya:2016ecj,Bhar:2017ynp,Ovalle:2017fgl,Gabbanelli:2018bhs,Ovalle:2017wqi,Maurya:2015maa,Abellan:2020nkl}.  A formulation of the TOV equations that is particularly useful for developing a gauge-invariant perturbation theory is the one given in \cite{Carloni:2017rpu,Carloni:2017bck}. The advantage of such a formulation is twofold: it is manifestly covariant, and the introduction of anisotropic stress is immediate with an appropriate choice of frame. The reason behind these features is the underlying formalism on which that version of the TOV equations is based: the 1+1+2 covariant approach \cite{Clarkson_Barrett_2003,Clarkson_2007}. The 1+1+2 covariant formalism can be viewed as a semi-tetradic approach to relativistic gravitation, where the Ricci and Bianchi identities are combined with the gravitational field equations and decomposed into equations for a set of scalars, vectors, and tensors defined on a two-dimensional surface. These quantities intuitively describe both the geometry and the thermodynamics of a given spacetime, without the need to introduce coordinates.
The 1+1+2 formalism simplifies considerably for spacetimes with local rotational symmetries, the so-called {\it LRS spacetimes} \cite{Ellis:1966ta,Ellis:1968vb,Stewart:1967tz,MacCallum:2022lwl}. More specifically, using the 1+1+2 formalism, LRS spacetimes can be completely characterized only in terms of Lorentz scalars. Indeed, LRS spacetimes are particularly suitable for characterizing non-rotating or slowly rotating compact stellar objects, making the 1+1+2 approach especially useful in such contexts.

However, the real power of the 1+1+2 formalism becomes apparent when studying linear perturbations.  Using this covariant formalism, it is possible to construct a gauge-invariant description of the evolution of these perturbations. For instance, the 1+1+2 formalism has been shown, since its first proposal, to reproduce the key equations of black hole perturbations \cite{Clarkson_Barrett_2003}, and it has also been used to treat scalar field and electromagnetic perturbations~\cite{Betschart_Clarkson_2004} on LRS spacetimes. Much later,  this work was extended to perturbed cosmological  spacetimes~\cite{Tornkvist:2019wwy,Semren:2022jmq}. In Ref.~\cite{Luz_Carloni_2024a}, we developed a complete theory of perturbation for LRS class II spacetimes associated with compact stellar objects, permeated by a general matter fluid. In the two following papers~\cite{Luz_Carloni_2024b,Luz_Carloni_2024c}, we focused on the case of isotropic sources, examining the perturbation in two frames: one comoving with matter (the comoving frame), and one at rest with respect to an observer at spatial infinity (the static frame).  We showed that, provided the background and perturbed equation of state are sufficiently regular, both sets of perturbation equations can also be integrated exactly per power series, other than numerically.  In general, we found that although the results in the two frames are compatible, the integration of the perturbation system is, in fact, much easier in the static frame at the cost of having a more complicated thermodynamic description of the matter fluid. By imposing dynamical stability, we also obtained a stronger constraint on the maximum compactness of isotropic objects, surpassing that of the Buchdahl theorem.

This article aims to extend the results from the last three papers to the case of self-gravitating fluids with anisotropic pressure in the comoving frame.  One common difficulty in developing a framework for perturbations of anisotropic fluids is that the details of the model for anisotropic stresses strongly influence the description and behavior of the perturbations, making it difficult to derive general results. Nonetheless, we address this problem by introducing a general function to characterize the anisotropic ansatz, which can readily incorporate all known models of anisotropic stresses and facilitates comparison of their differences.  Such a setup is then used to determine the behavior of the perturbations of anisotropic backgrounds or isotropic backgrounds that experience anisotropic perturbations. We also argue that, with a suitable choice of background, one can determine a maximum compactness also in the case of anisotropic stars.

The paper is organized as follows. In Section~\ref{Sec:Perturbation equations}, we present the general covariant gauge invariant perturbation equations in the 1+1+2 formalism for compact stellar models in the presence of anisotropic stresses. In Section~\ref{Sec:Anisotropic_ansatze}, we consider perturbations assuming the anisotropic ansatze resultant from various non-equilibrium thermodynamic theories, using a unified formulation based on the 1+1+2 covariant variables. In Section~\ref{Sec:Effects_shear_relaxation}, we discuss the effects of shear viscosity and relaxational time on the evolution of perturbations. In Section~\ref{Sec:Strange_stars}, we study the eigenmodes of perturbation of strange stars.
Section~\ref{Sec:Maximum_compactness} is dedicated to the derivation of a bound for the maximum compactness of an anisotropic stellar object. Finally, Section~\ref{Sec:Conclusion} presents the conclusions. This article has two appendices.
In Appendix~\ref{Appendix_sec:EoS2}, we characterize popular anisotropic ansatze as particular cases of the general unifying formulation. In Appendix~\ref{Appendix_sec:Auxiliary_functions_comoving_frame}, we present some auxiliary quantities that characterize the perturbations.

Throughout the article, we will consider the metric signature $(-+++)$ and, except in Section~\ref{Sec:Strange_stars}, we work in the geometrized unit system
where $8\pi G=c=1$.

\section{Perturbation equations} \label{Sec:Perturbation equations}

\subsection{The equilibrium spacetime and the gauge-invariant perturbation variables}

Our purpose is to develop a general formulation of the field equations
that characterize linear perturbations of static, isotropic, non-dissipative
self-gravitating fluids within the theory of General Relativity.

Concretely, we start from the Einstein field equations
\begin{equation}
R_{\alpha\beta}-\frac{1}{2}g_{\alpha\beta}R+g_{\alpha\beta}\Lambda=T_{\alpha\beta}\,,
\end{equation}
where $R_{\alpha\beta}$ are the components of the Ricci tensor in
some local coordinate system, $R$ the Ricci scalar, $g_{\alpha\beta}$
the components of the metric tensor, $\Lambda$, the cosmological constant
and $T_{\alpha\beta}$ the components of the stress-energy tensor
of the matter fields. 

The fluid source of a static, spherically symmetric solution of the field equations can be generically characterized by a stress-energy tensor of the form
\begin{equation}
T_{\alpha\beta}=\left(\mu+p\right)u_{\alpha}u_{\beta}+pg_{\alpha\beta}+\pi_{\alpha\beta}\,,
\end{equation}
where $u$ is the 4-velocity of an observer locally comoving with
the volume elements of the fluid, $\mu$ is the energy density, $p$
the isotropic pressure of the matter fluid and $\pi_{\alpha\beta}$
characterize the anisotropic stresses within the fluid. Introducing
a spacelike vector field, $e$, such that $u_{\alpha}e^{\alpha}=0$,
$\pi_{\alpha\beta}$ can be decomposed in scalar, $\Pi$, vector,
$\Pi_{\alpha}$, and tensor, $\Pi_{\alpha\beta}$, components with
respect to $e$. This is the essence of the semi-tetradic 1+1+2 formalism~\citep{Clarkson_Barrett_2003,Betschart_Clarkson_2004,Clarkson_2007}.
For a static, spherically symmetric spacetime, $e$ can be chosen
in such a way that the most general stress-energy tensor can be written as
\begin{equation}
T_{\alpha\beta}=\left(\mu+p-\frac{1}{2}\Pi\right)u_{\alpha}u_{\beta}+\frac{3}{2}\Pi e_{\alpha}e_{\beta}+\left(p-\frac{1}{2}\Pi\right)g_{\alpha\beta}\,.\label{Comoving_Perturbation_eqs:Stress_energy_tensor_general}
\end{equation}
The isotropic and anistropic pressures, $p$ and $\Pi$, respectively,
are related to the radial and tangential pressures by Eq.~\eqref{Appendix_ansatze_eq:Radial_tangential_pressures_p_Pi_relations}.
For a perfect fluid, $\Pi=0$.

Now, we will consider the equilibrium spacetime to be composed of
two solutions of the Einstein field equations. An interior static,
spatially compact, spherically symmetric solution with a matter source
characterized by~\eqref{Comoving_Perturbation_eqs:Stress_energy_tensor_general},
and an exterior, vacuum solution, given by a regular branch of the
Schwarzschild spacetime. The two solutions are smoothly matched at
a common timelike hypersurface with constant circumferential radius
$r=r_{b}$. We will refer to such solutions as {\it stars}.

In what follows, it is useful to derive the perturbation equations
using a covariant formalism. However, in Sec.~\ref{subsec:Breaking_covariance}
and onward, we will explicitly adopt the Schwarzschild coordinate system
defined by a radially static observer at spatial infinity. In this
coordinates, the equilibrium spacetime can then be characterized by
a line element of the form
\begin{equation}
ds^{2}=-g_{tt}\,dt^{2}+g_{rr}\,dr^{2}+r^{2}d\Omega^{2}\,,\label{Comoving_Perturbation_eqs:general_static_line_element}
\end{equation}
where the metric coefficients $g_{tt}$ and $g_{rr}$ are assumed
to be functions of the radial coordinate $r$, and $d\Omega^{2}$ is
the line element of the unit 2-sphere. Since we will consider both the equilibrium and the perturbed spacetimes, from hereon, we will use a subscript ``0'' to refer to quantities that characterize the
equilibrium spacetime. From Eq.~\eqref{Comoving_Perturbation_eqs:general_static_line_element}, we can define the scalar functions 
\begin{equation}
\begin{aligned}\phi_{0} & =\frac{2}{r\sqrt{\left(g_{0}\right)_{rr}}}\,,\\
\mathcal{A}_{0} & =\frac{1}{2\left(g_{0}\right)_{tt}\sqrt{\left(g_{0}\right)_{rr}}}\frac{d\left(g_{0}\right)_{tt}}{dr}\,,\\
\mathcal{E}_{0} & =\frac{1}{3}\mu_{0}+p_{0}-\frac{2}{3}\Lambda-\mathcal{A}_{0}\phi_{0}+\frac{1}{2}\Pi_{0}\,,
\end{aligned}
\label{Comoving_Perturbation_eqs:Background_phi_A_E}
\end{equation}
where $\phi_{0}$ characterizes the spatial expansion of the normalized
radial gradient vector field, $\mathcal{A}_{0}$ is the radial component
of the 4-acceleration of an observer locally comoving with the volume
elements of the matter fluid, and $\mathcal{E}_{0}$ represents the
nontrivial component of the Weyl tensor, characterizing tidal forces.

Following the Stewart-Walker Lemma~\citep{Stewart_Walker_1974}, we will describe the perturbed spacetime using variables that are
manifestly gauge invariant, that is, we will consider variables that
vanish identically in the equilibrium spacetime and the cosmological
constant, $\Lambda$, which is identical in both spacetimes. Namely,
let
\begin{equation}
\begin{aligned}\mathsf{m} & :=\dot{\mu}\,, & \mathsf{p} & :=\dot{p}\,, & \mathcal{P} & :=\dot{\Pi}\,, & \mathsf{A} & :=\dot{\mathcal{A}}\,, & \mathsf{F} & :=\dot{\phi}\,, & \mathsf{E} & :=\dot{\mathcal{E}}\,,\end{aligned}
\label{Comoving_Perturbation_eqs:GI_dot_derivatives_definition}
\end{equation}
represent the proper time derivatives of the various scalar quantities
that can be used to describe the spacetime. The equilibrium spacetime
is assumed to be static. Therefore, the quantities in Eq.~\eqref{Comoving_Perturbation_eqs:GI_dot_derivatives_definition}
are identically zero in the background. In addition to these variables,
to fully characterize the perturbed spacetime and matter fields, we
will also consider the expansion scalar $\theta$ associated with
the integral curves of the $u$ vector field and the scalar component,
with respect to $e$, of the shear tensor, $\Sigma$ (see Refs.~\citep{Luz_Carloni_2024a,Luz_Carloni_2024b,Luz_Carloni_2024c}
for details). Both of these quantities also vanish identically
in the static background. The set $\left\{ \mathsf{m},\mathsf{p},\mathcal{P},\Lambda,\mathsf{A},\mathsf{F},\mathsf{E},\theta,\Sigma\right\} $
contains only gauge invariant variables and can fully characterize
the linear, radial, adiabatic perturbations of stars.

\subsection{Harmonic decomposition\label{subsec:Harmonic-decomposition}}

In Ref.~\citep{Luz_Carloni_2024a}, we have found the linearized Einstein
field equations considering the 1+1+2 covariant formalism. Although the linearized system is much simpler than the fully fledged set of
field equations, it is still a system of partial differential equations,
too complicated to solve, given a generic background spacetime. On the other hand, since the equilibrium spacetime is assumed to be static
and spherically symmetric, we can make use of the exact symmetries
to harmonically decompose the perturbation variables and find the
associated system of ordinary differential equations for the harmonic
coefficients.

Let $\chi$ be a first-order, gauge-invariant scalar quantity. Given
the eigenfunctions $e^{i\upsilon\tau}$ of the Laplace operator in
$\mathbb{R}$ and the Spherical Harmonics, $Y_{\ell m}$, we can write
the harmonic decomposition
\begin{equation}
\chi=\sum_{\upsilon}\left(\sum_{\ell=0}^{+\infty}\sum_{m=-\ell}^{\ell}\Psi_{\chi}^{\left(\upsilon,\ell\right)}Y_{\ell m}\right)e^{i\upsilon\tau}\,,\label{Comoving_Perturbation_eqs:general_harmonic_expansion}
\end{equation}
where $\tau$ is the proper time of the comoving observer in the background
spacetime and $\upsilon$ are the associated eigenfrequencies. Depending
on the boundary conditions of the problem, the symbol $\sum_{\upsilon}$
can stand for a discrete sum or an integral in $\upsilon$, and the
coefficients $\Psi_{\chi}^{\left(\upsilon,\ell\right)}$ are functions
of $r$.

In this article, we are interested in studying the properties of isotropic
perturbations, therefore we can disregard all multipoles with $\ell\geq1$
and consider only the scalar modes, with $\ell=0$, that is, we will
consider that the only non-vanishing coefficients of the harmonic
expansion~\eqref{Comoving_Perturbation_eqs:general_harmonic_expansion}
are $\Psi_{\chi}^{\left(\upsilon,0\right)}$. To lighten the notation, from here on out, we
will drop the subscript for $\ell$. Then, in the considered setup,
any first-order, gauge-invariant scalar quantity, $\chi$, can be
decomposed as
\begin{equation}
\chi=\sum_{\upsilon}\Psi_{\chi}^{\left(\upsilon\right)}\left(r\right)Y_{00}\,e^{i\upsilon\tau}\,.\label{Comoving_Perturbation_eqs:Radial_Adiabatic_general_harmonic_decompostion_upsilon}
\end{equation}

\subsection{Linearized field equations in the comoving frame}

In the comoving frame, i.e., in the absence of 
heat flows, the equations for the radial perturbations read~\citep{Luz_Carloni_2024a}
\begin{equation}
\begin{aligned}\widehat{\Psi}_{\mathsf{A}}^{\left(\upsilon\right)} & =\left(\frac{1}{2}\mu_{0}+3p_{0}-\frac{5}{2}\Lambda-3\mathcal{A}_{0}\phi_{0}-\frac{3}{2}\mathcal{A}_{0}^{2}+\frac{3}{2}\Pi_{0}-\frac{3}{2}\upsilon^{2}\right)\Psi_{\Sigma}^{\left(\upsilon\right)}\\
 & -\left(\mu_{0}-3p_{0}+4\Lambda+6\mathcal{A}_{0}\phi_{0}-6\Pi_{0}\right)\frac{\Pi_{0}}{4\left(\mu_{0}+p_{0}\right)}\Psi_{\Sigma}^{\left(\upsilon\right)}\\
 & -\left(3\mathcal{A}_{0}-\frac{1}{2}\phi_{0}\right)\Psi_{\mathsf{A}}^{\left(\upsilon\right)}-\frac{3}{2}\Psi_{\mathcal{P}}^{\left(\upsilon\right)}+\frac{2\mathcal{E}_{0}+\Pi_{0}}{2\left(\mu_{0}+p_{0}\right)}\Psi_{\mathsf{m}}^{\left(\upsilon\right)}\,,
\end{aligned}
\label{Comoving_Perturbation_eqs:Psi_A_hat}
\end{equation}
\begin{equation}
\begin{aligned}\widehat{\Psi}_{\mathsf{p}}^{\left(\upsilon\right)}+\widehat{\Psi}_{\mathcal{P}}^{\left(\upsilon\right)} & =\left[-\frac{9}{4}\Pi_{0}\phi_{0}-\left(\mu_{0}+p_{0}-\Pi_{0}\right)\mathcal{A}_{0}+\frac{2\Pi_{0}^{2}\mathcal{A}_{0}}{\mu_{0}+p_{0}}\right]\Psi_{\Sigma}^{\left(\upsilon\right)}-2\mathcal{A}_{0}\Psi_{\mathsf{p}}^{\left(\upsilon\right)}\\
 & +\frac{2}{3}\left(\frac{2\Pi_{0}}{\mu_{0}+p_{0}}-1\right)\mathcal{A}_{0}\Psi_{\mathsf{m}}^{\left(\upsilon\right)}-\left(\mu_{0}+p_{0}+\Pi_{0}\right)\Psi_{\mathsf{A}}^{\left(\upsilon\right)}-\left(\frac{3}{2}\phi_{0}+2\mathcal{A}_{0}\right)\Psi_{\mathcal{P}}^{\left(\upsilon\right)}\,,
\end{aligned}
\end{equation}
\begin{equation}
\frac{2}{3\left(\mu_{0}+p_{0}\right)}\widehat{\Psi}_{\mathsf{m}}^{\left(\upsilon\right)}+\left(1+\frac{\Pi_{0}}{\mu_{0}+p_{0}}\right)\widehat{\Psi}_{\Sigma}^{\left(\upsilon\right)}=\frac{2\left(\widehat{\mu}_{0}+\widehat{p}_{0}\right)}{3\left(\mu_{0}+p_{0}\right)^{2}}\Psi_{\mathsf{m}}^{\left(\upsilon\right)}-\left[\frac{3}{2}\phi_{0}+\frac{\widehat{\Pi}_{0}}{\mu_{0}+p_{0}}-\frac{\Pi_{0}\left(\widehat{\mu}_{0}+\widehat{p}_{0}\right)}{\left(\mu_{0}+p_{0}\right)^{2}}\right]\Psi_{\Sigma}^{\left(\upsilon\right)}\,,\label{Comoving_Perturbation_eqs:Psi_sigma_mDot_hat:}
\end{equation}
with the constraint equations
\begin{equation}
\left(\Lambda+\mathcal{A}_{0}\phi_{0}+\mathcal{A}_{0}^{2}-p_{0}-\Pi_{0}+\upsilon^{2}\right)\left[\frac{2}{3\left(\mu_{0}+p_{0}\right)}\Psi_{\mathsf{m}}^{\left(\upsilon\right)}+\left(1+\frac{\Pi_{0}}{\mu_{0}+p_{0}}\right)\Psi_{\Sigma}^{\left(\upsilon\right)}\right]=\phi_{0}\Psi_{\mathsf{A}}^{\left(\upsilon\right)}-\Psi_{\mathsf{p}}^{\left(\upsilon\right)}-\Psi_{\mathcal{P}}^{\left(\upsilon\right)}\,,\label{Comoving_Perturbation_eqs:Constraint_upsilon}
\end{equation}
\begin{equation}
\left(\mu_{0}+p_{0}\right)\Psi_{\theta}^{\left(\upsilon\right)}=-\frac{3}{2}\Pi_{0}\Psi_{\Sigma}^{\left(\upsilon\right)}-\Psi_{\mathsf{m}}^{\left(\upsilon\right)}\,,
\end{equation}
\begin{equation}
\Psi_{\mathsf{F}}^{\left(\upsilon\right)}=-\frac{2\mathcal{A}_{0}-\phi_{0}}{3\left(\mu_{0}+p_{0}\right)}\Psi_{\mathsf{m}}^{\left(\upsilon\right)}-\left(\mathcal{A}_{0}-\frac{1}{2}\phi_{0}\right)\left(1+\frac{\Pi_{0}}{\mu_{0}+p_{0}}\right)\Psi_{\Sigma}^{\left(\upsilon\right)}\,,
\end{equation}
\begin{equation}
\Psi_{\mathsf{E}}^{\left(\upsilon\right)}=-\frac{1}{2}\left(\mu_{0}+p_{0}-\frac{1}{2}\Pi_{0}-3\mathcal{E}_{0}\right)\left(1+\frac{\Pi_{0}}{\mu_{0}+p_{0}}\right)\Psi_{\Sigma}^{\left(\upsilon\right)}+\frac{\Pi_{0}+6\mathcal{E}_{0}}{6\left(\mu_{0}+p_{0}\right)}\Psi_{\mathsf{m}}^{\left(\upsilon\right)}-\frac{1}{2}\Psi_{\mathcal{P}}^{\left(\upsilon\right)}\,,\label{Comoving_Perturbation_eqs:Constraint_Curly_E}
\end{equation}
where the hat represents the directional derivative with respect to
the vector field $e$, that is, for a generic variable $\chi$, $\widehat{\chi}:=\nabla_{e}\chi$.
We remark that Eq.~\eqref{Comoving_Perturbation_eqs:Constraint_upsilon}
cannot, in general, be used to reduce the system of evolution equations to two
equations, since this equation is not propagated. Therefore, for a specific matter model and imposing boundary conditions, solutions of the linearized Einstein field equations
follow from solving the differential equations together with the constraint
equation~\eqref{Comoving_Perturbation_eqs:Constraint_upsilon}. 
Nonetheless, it is straightforward to show that a solution of the differential equations~\eqref{Comoving_Perturbation_eqs:Psi_A_hat}--\eqref{Comoving_Perturbation_eqs:Psi_sigma_mDot_hat:} verifies
Eq.\,\eqref{Comoving_Perturbation_eqs:Constraint_upsilon} if, and only if,
\begin{equation}\label{eqUps}
\upsilon\left(\widehat{\upsilon}+\mathcal{A}_{0}\upsilon\right)=0\,.
\end{equation}
This equation can be readily integrated for $\upsilon$.
For instance, breaking covariance, assuming an asymptotically flat
spacetime, and considering the Schwarzschild coordinates defined by
a radially static observer at spatial infinity, we find
\begin{equation}
\upsilon\left(r\right)=\lambda\exp\left(-\int\frac{2\mathcal{A}_{0}}{r\phi_{0}}dr\right)=\lambda\sqrt{g_{0}^{tt}}\,,
\label{eq:Ups_radial}
\end{equation}
where $\lambda$ represents the constant, eigenfrequency measured
by the static observer. Given the solutions for $\upsilon$ in some coordinate system, one can solve the full perturbation system
simply by substituting the expression for $\upsilon$ in Eqs.~\eqref{Comoving_Perturbation_eqs:Psi_A_hat}--\eqref{Comoving_Perturbation_eqs:Psi_sigma_mDot_hat:},
provided the differential system is closed with appropriate boundary conditions and equations characterizing the matter fields.

\subsection{A general equation of state}

The system~\eqref{Comoving_Perturbation_eqs:Psi_A_hat}--\eqref{Comoving_Perturbation_eqs:Constraint_Curly_E}
does not close since it requires the specification of the matter model.
In the case of radial adiabatic perturbations, the matter model can
be constrained by providing a single equation of state relating the
energy density with the isotropic pressure or by imposing one of the
matter thermodynamic quantities to be some known function of the spacetime.
However, for an anisotropic fluid, it is necessary to provide equations
that relate the three matter variables that characterize the perturbed
fluid: $\mu$, $p$ and $\Pi$.

In that regard, consider a general equation of state in the comoving
frame
\begin{equation}
f\left(\mu,p,\Pi\right)=0\,,
\end{equation}
where $f$ is assumed to be a differentiable function. We remark that
$f$ does not necessarily need to be related to the equation of
state of the unperturbed fluid. Taking the derivative and making use
of the Chain Rule yields
\begin{equation}
\mathsf{m}f_{\mu}\left(\mu,p,\Pi\right)+\mathsf{p}f_{p}\left(\mu,p,\Pi\right)+\mathcal{P}f_{\Pi}\left(\mu,p,\Pi\right)=0\,,
\end{equation}
where $f_{\mu}=\partial_{\mu}f$, $f_{p}=\partial_{p}f$ and $f_{\Pi}=\partial_{\Pi}f$
represent partial derivatives of $f$. Then, at linear order, we have
\begin{equation}
\mathsf{m}f_{\mu}\left(\mu_{0},p_{0},\Pi_{0}\right)+\mathsf{p}f_{p}\left(\mu_{0},p_{0},\Pi_{0}\right)+\mathcal{P}f_{\Pi}\left(\mu_{0},p_{0},\Pi\right)=0\,.\label{Comoving_Perturbation_eqs:EoS_general}
\end{equation}

To not saturate the notation, in what follows, we will
drop the arguments of the derivatives, being implicit from the last
equation that these depend on the background thermodynamic variables.
Assuming that $f_{\mu}$ does not vanish in the interior of the perturbed
star, we can use Eq.~\eqref{Comoving_Perturbation_eqs:EoS_general}
to remove the $\Psi_{\mathsf{m}}$ coefficient in Eqs.~\eqref{Comoving_Perturbation_eqs:Psi_A_hat}--\eqref{Comoving_Perturbation_eqs:Constraint_Curly_E},
such that 
\begin{equation}
\begin{aligned}\widehat{\Psi}_{\mathsf{A}}^{\left(\upsilon\right)} & =\left(\frac{1}{2}\mu_{0}+3p_{0}-\frac{5}{2}\Lambda-3\mathcal{A}_{0}\phi_{0}-\frac{3}{2}\mathcal{A}_{0}^{2}+\frac{3}{2}\Pi_{0}-\frac{3}{2}\upsilon^{2}\right)\Psi_{\Sigma}^{\left(\upsilon\right)}\\
 & -\left(\mu_{0}-3p_{0}+4\Lambda+6\mathcal{A}_{0}\phi_{0}-6\Pi_{0}\right)\frac{\Pi_{0}}{4\left(\mu_{0}+p_{0}\right)}\Psi_{\Sigma}^{\left(\upsilon\right)}\\
 & -\left(3\mathcal{A}_{0}-\frac{1}{2}\phi_{0}\right)\Psi_{\mathsf{A}}^{\left(\upsilon\right)}-\frac{\left(2\mathcal{E}_{0}+\Pi_{0}\right)f_{p}}{2\left(\mu_{0}+p_{0}\right)f_{\mu}}\Psi_{\mathsf{p}}^{\left(\upsilon\right)}-\left[\frac{\left(2\mathcal{E}_{0}+\Pi_{0}\right)f_{\Pi}}{2\left(\mu_{0}+p_{0}\right)f_{\mu}}+\frac{3}{2}\right]\Psi_{\mathcal{P}}^{\left(\upsilon\right)}\,,
\end{aligned}
\label{Comoving_Perturbation_eqs:A_hat_after_EOS}
\end{equation}
\begin{equation}
\begin{aligned}\widehat{\Psi}_{\mathsf{p}}^{\left(\upsilon\right)}+\widehat{\Psi}_{\mathcal{P}}^{\left(\upsilon\right)} & =\left[-\frac{9}{4}\Pi_{0}\phi_{0}-\left(\mu_{0}+p_{0}-\Pi_{0}\right)\mathcal{A}_{0}+\frac{2\Pi_{0}^{2}\mathcal{A}_{0}}{\mu_{0}+p_{0}}\right]\Psi_{\Sigma}^{\left(\upsilon\right)}-\left(\mu_{0}+p_{0}+\Pi_{0}\right)\Psi_{\mathsf{A}}^{\left(\upsilon\right)}\\
 & -\left[\frac{2}{3}\left(\frac{2\Pi_{0}}{\mu_{0}+p_{0}}-1\right)\frac{f_{\Pi}}{f_{\mu}}\mathcal{A}_{0}+\left(\frac{3}{2}\phi_{0}+2\mathcal{A}_{0}\right)\right]\Psi_{\mathcal{P}}^{\left(\upsilon\right)}\\
 & -\left[2\mathcal{A}_{0}+\frac{2}{3}\left(\frac{2\Pi_{0}}{\mu_{0}+p_{0}}-1\right)\frac{f_{p}}{f_{\mu}}\mathcal{A}_{0}\right]\Psi_{\mathsf{p}}^{\left(\upsilon\right)}\,,
\end{aligned}
\end{equation}
\begin{equation}
\begin{aligned}\left(\mu_{0}+p_{0}+\Pi_{0}\right)\widehat{\Psi}_{\Sigma}^{\left(\upsilon\right)}-\frac{2}{3f_{\mu}}\left[f_{p}\widehat{\Psi}_{\mathsf{p}}^{\left(\upsilon\right)}+f_{\Pi}\widehat{\Psi}_{\mathcal{P}}^{\left(\upsilon\right)}\right] & =\frac{2}{3}\left[D_{e}\left(\frac{f_{p}}{f_{\mu}}\right)\Psi_{\mathsf{p}}^{\left(\upsilon\right)}+D_{e}\left(\frac{f_{\Pi}}{f_{\mu}}\right)\Psi_{\mathcal{P}}^{\left(\upsilon\right)}\right]\\
 & -\left[\frac{3}{2}\left(\mu_{0}+p_{0}\right)\phi_{0}+\widehat{\Pi}_{0}-\frac{\Pi_{0}\left(\widehat{\mu}_{0}+\widehat{p}_{0}\right)}{\mu_{0}+p_{0}}\right]\Psi_{\Sigma}^{\left(\upsilon\right)}\\
 & -\frac{2\left(\widehat{\mu}_{0}+\widehat{p}_{0}\right)}{3\left(\mu_{0}+p_{0}\right)f_{\mu}}\left(f_{p}\Psi_{\mathsf{p}}^{\left(\upsilon\right)}+f_{\Pi}\Psi_{\mathcal{P}}^{\left(\upsilon\right)}\right)\,,
\end{aligned}
\end{equation}
and the constraints 
\begin{equation}
\begin{aligned}
	\left(\mathcal{U}+\upsilon^{2}\right)\left(1+\frac{\Pi_{0}}{\mu_{0}+p_{0}}\right)\Psi_{\Sigma}^{\left(\upsilon\right)} & =\phi_{0}\Psi_{\mathsf{A}}^{\left(\upsilon\right)}+\left[\frac{2\left(\Lambda+\mathcal{A}_{0}\phi_{0}+\mathcal{A}_{0}^{2}-p_{0}-\Pi_{0}+\upsilon^{2}\right)f_{p}}{3\left(\mu_{0}+p_{0}\right)f_{\mu}}-1\right]\Psi_{\mathsf{p}}^{\left(\upsilon\right)}\\
 & +\left[\frac{2\left(\Lambda+\mathcal{A}_{0}\phi_{0}+\mathcal{A}_{0}^{2}-p_{0}-\Pi_{0}+\upsilon^{2}\right)f_{\Pi}}{3\left(\mu_{0}+p_{0}\right)f_{\mu}}-1\right]\Psi_{\mathcal{P}}^{\left(\upsilon\right)}\,,
\end{aligned}
\label{Comoving_Perturbation_eqs:Extra_constraint_after_EOS}
\end{equation}
\begin{equation}
\Psi_{\mathsf{F}}^{\left(\upsilon\right)}=\frac{2\mathcal{A}_{0}-\phi_{0}}{3\left(\mu_{0}+p_{0}\right)f_{\mu}}\left(f_{p}\Psi_{\mathsf{p}}^{\left(\upsilon\right)}+f_{\Pi}\Psi_{\mathcal{P}}^{\left(\upsilon\right)}\right)-\left(\mathcal{A}_{0}-\frac{1}{2}\phi_{0}\right)\left(1+\frac{\Pi_{0}}{\mu_{0}+p_{0}}\right)\Psi_{\Sigma}^{\left(\upsilon\right)}\,,
\end{equation}
\begin{equation}
\begin{aligned}\Psi_{\mathsf{E}}^{\left(\upsilon\right)} & =-\frac{1}{2}\left(\mu_{0}+p_{0}-\frac{1}{2}\Pi_{0}-3\mathcal{E}_{0}\right)\left(1+\frac{\Pi_{0}}{\mu_{0}+p_{0}}\right)\Psi_{\Sigma}^{\left(\upsilon\right)}-\frac{\left(\Pi_{0}+6\mathcal{E}_{0}\right)f_{p}}{6\left(\mu_{0}+p_{0}\right)f_{\mu}}\Psi_{\mathsf{p}}^{\left(\upsilon\right)}\\
 & -\left[\frac{1}{2}+\frac{\left(\Pi_{0}+6\mathcal{E}_{0}\right)f_{\Pi}}{6\left(\mu_{0}+p_{0}\right)f_{\mu}}\right]\Psi_{\mathcal{P}}^{\left(\upsilon\right)}\,,
\end{aligned}
\label{Comoving_Perturbation_eqs:E_after_EOS}
\end{equation}
\begin{equation}
\left(\mu_{0}+p_{0}\right)\Psi_{\theta}^{\left(\upsilon\right)}=-\frac{3}{2}\Pi_{0}\Psi_{\Sigma}^{\left(\upsilon\right)}+\frac{f_{p}}{f_{\mu}}\Psi_{\mathsf{p}}^{\left(\upsilon\right)}+\frac{f_{\Pi}}{f_{\mu}}\Psi_{\mathcal{P}}^{\left(\upsilon\right)}\,,\label{Comoving_Perturbation_eqs:Theta_after_EOS}
\end{equation}
\begin{equation}
f_{\mu}\Psi_{\mathsf{m}}^{\left(\upsilon\right)}+f_{p}\Psi_{\mathsf{p}}^{\left(\upsilon\right)}+f_{\Pi}\Psi_{\mathcal{P}}^{\left(\upsilon\right)}=0\,,\label{Comoving_Perturbation_eqs:Mu_after_EOS}
\end{equation}
where
\begin{equation}
\mathcal{U}=\Lambda+\mathcal{A}_{0}\phi_{0}+\mathcal{A}_{0}^{2}-p_{0}-\Pi_{0}\,.
\end{equation}
From this system, it is clear that a further relation is needed to describe how the anisotropic stresses are generated in the perturbed
fluid.

\subsection{Completing the matter model: the anisotropic ansatz\label{subsec:Comoving_complete_matter_model}}

An equation of state, generically written in the form of Eq.~\eqref{Comoving_Perturbation_eqs:EoS_general},
only partially characterizes the matter fields permeating spacetime.
To completely specify the matter model, we need to provide a relation
that connects the anisotropic stresses within the fluid with the remaining
matter variables and the kinematical quantities of the congruence
associated with the paths of the elements of volume. In the simplest case, we may consider that those variables are algebraically related
by a single equation of the form
\begin{equation}
g\left(\mu,p,\Pi,\mathcal{A},\phi,\Sigma,\theta\right)=0\,.\label{Comoving_Perturbation_eqs:EoS_2_algebraic_general}
\end{equation}
where $g$ is a differentiable function of its arguments. Indeed, various popular models presented in the literature can be cast in
the form of Eq.~\eqref{Comoving_Perturbation_eqs:EoS_2_algebraic_general},
yielding particular realizations of this equation (cf. Appendix~\ref{Appendix_sec:EoS2}).

Equation~\eqref{Comoving_Perturbation_eqs:EoS_2_algebraic_general}
directly relates the matter variables to the kinematical quantities.
In the context of perturbation theory, this equation might not be
gauge invariant since it might relate quantities that do not vanish
in the unperturbed spacetime. However, Eq.~\eqref{Comoving_Perturbation_eqs:EoS_2_algebraic_general}
implies the gauge invariant equation
\begin{equation}
\left(\partial_{\mu}g\right)\mathsf{m}+\left(\partial_{p}g\right)\mathsf{p}+\left(\partial_{\Pi}g\right)\mathcal{P}+\left(\partial_{\mathcal{A}}g\right)\mathsf{A}+\left(\partial_{\phi}g\right)\mathsf{F}+\left(\partial_{\Sigma}g\right)\dot{\Sigma}+\left(\partial_{\theta}g\right)\dot{\theta}=0\,,\label{Comoving_Perturbation_eqs:EoS_2_algebric_general_linear_derivatives}
\end{equation}
where the symbol $\partial$ represents partial derivatives of $g$ with respect to the indicated variable, and, for simplicity's sake, we omit the arguments of the derivatives, being implicit that, in the context of linear perturbation theory, those only depend on the background thermodynamic variables.

Equation~\eqref{Comoving_Perturbation_eqs:EoS_2_algebric_general_linear_derivatives}
provides a unifying framework to treat perturbations of self-gravitating
fluids provided an algebraic ansatz for the relation between the anisotropic
pressure and the remaining variables. However, some thermodynamic
models lead to differential equations for the components of the pressure.
Notably, this follows in the context of causal non-equilibrium thermodynamics~\cite{Israel_1976,Israel_Stewart_1979}.
In the case when the relations represent differential equations in
time, Eq.~\eqref{Comoving_Perturbation_eqs:EoS_2_algebraic_general}
can be generalized to explicitly allow the presence of derivatives
of any order, that is
\begin{equation}
g\left(\mu,p,\Pi,\mathcal{A},\phi,\Sigma,\theta,\dot{\mu},\dot{p},\dot{\Pi},\dot{\mathcal{A}},\dot{\phi},\dot{\Sigma},\dot{\theta},...\right)=0\,.\label{Comoving_Perturbation_eqs:EoS_2_differential_general_full}
\end{equation}
Upon differentiation of Eq.~\eqref{Comoving_Perturbation_eqs:EoS_2_differential_general_full},
we find the gauge-invariant equation
\begin{equation}
\begin{aligned}\left(\partial_{\mu}g\right)\mathsf{m}+\left(\partial_{p}g\right)\mathsf{p}+\left(\partial_{\Pi}g\right)\mathcal{P}+\left(\partial_{\mathcal{A}}g\right)\mathsf{A}+\left(\partial_{\phi}g\right)\mathsf{F}+\left(\partial_{\Sigma}g\right)\dot{\Sigma}+\left(\partial_{\theta}g\right)\dot{\theta}\\
+\left(\partial_{\dot{\mu}}g\right)\dot{\mathsf{m}}+\left(\partial_{\dot{p}}g\right)\dot{\mathsf{p}}+\left(\partial_{\dot{\Pi}}g\right)\dot{\mathcal{P}}+\left(\partial_{\dot{\mathcal{A}}}g\right)\dot{\mathsf{A}}+\left(\partial_{\dot{\phi}}g\right)\dot{\mathsf{F}}+\left(\partial_{\dot{\Sigma}}g\right)\ddot{\Sigma}+\left(\partial_{\dot{\theta}}g\right)\ddot{\theta}+... & =0
\end{aligned}
\label{Comoving_Perturbation_eqs:EoS_2_differential_general_linear_derivatives}
\end{equation}

Applying the harmonic decomposition of Sec.~\ref{subsec:Harmonic-decomposition},
we can decompose Eq.~\eqref{Comoving_Perturbation_eqs:EoS_2_differential_general_linear_derivatives}
to find an algebraic equation for the harmonic coefficients $\left\{ \Psi_{\mathsf{m}}^{\left(\upsilon\right)},\Psi_{\mathsf{p}}^{\left(\upsilon\right)},\Psi_{\mathcal{P}}^{\left(\upsilon\right)},\Psi_{\mathsf{A}}^{\left(\upsilon\right)},\Psi_{\mathsf{F}}^{\left(\upsilon\right)},\Psi_{\Sigma}^{\left(\upsilon\right)},\Psi_{\theta}^{\left(\upsilon\right)}\right\} $.
Namely, Eq.~\eqref{Comoving_Perturbation_eqs:EoS_2_differential_general_linear_derivatives}
yields
\begin{equation}
g_{\mu}\Psi_{\mathsf{m}}^{\left(\upsilon\right)}+g_{p}\Psi_{\mathsf{p}}^{\left(\upsilon\right)}+g_{\Pi}\Psi_{\mathcal{P}}^{\left(\upsilon\right)}+g_{\mathcal{A}}\Psi_{\mathsf{A}}^{\left(\upsilon\right)}+g_{\phi}\Psi_{\mathsf{F}}^{\left(\upsilon\right)}+i\upsilon g_{\Sigma}\Psi_{\Sigma}^{\left(\upsilon\right)}+i\upsilon g_{\theta}\Psi_{\theta}^{\left(\upsilon\right)}=0\,.\label{Comoving_Perturbation_eqs:EoS_2_coeff}
\end{equation}
where, for a generic first-order variable $\chi$, the coefficients
\begin{equation}
g_{\chi}=\partial_{\chi}g+i\upsilon\partial_{\mathsf{\dot{\chi}}}g-\upsilon^{2}\partial_{\ddot{\chi}}g+...
\end{equation}

As we see, the case of having a differential equation characterizing the anisotropic stresses can be encapsulated in the form of
the coefficients $\left\{ g_{\mu},g_{p},g_{\Pi},g_{\mathcal{A}},g_{\phi},g_{\Sigma},g_{\theta}\right\} $.
In Appendix~\ref{Appendix_sec:EoS2}, we show that this is indeed
the case in the particular ansatz following from the Truncated
Israel-Stewart theory.

In the context of the linearized perturbation theory, many of the
1+1+2 potentials are related by algebraic constraint equations, Eqs.~\eqref{Comoving_Perturbation_eqs:Extra_constraint_after_EOS}--\eqref{Comoving_Perturbation_eqs:Theta_after_EOS}. Moreover,
imposing the equation of state~\eqref{Comoving_Perturbation_eqs:Mu_after_EOS},
one of the matter variables can be related to the remaining ones.
As such, in the setup considered in this article, to close the system~\eqref{Comoving_Perturbation_eqs:A_hat_after_EOS}--\eqref{Comoving_Perturbation_eqs:Mu_after_EOS},
the function $g$ can be considered to simply provide a relation of
the form
\begin{equation}
g\left(p,\Pi,\mathcal{A},\Sigma,\dot{p},\dot{\Pi},\dot{\mathcal{A}},\dot{\Sigma},...\right)=0\,,\label{Comoving_Perturbation_eqs:EoS_2_general_simpler}
\end{equation}
or, explicitly for the harmonic coefficients,
\begin{equation}
g_{p}\Psi_{\mathsf{p}}^{\left(\upsilon\right)}+g_{\Pi}\Psi_{\mathcal{P}}^{\left(\upsilon\right)}+g_{\mathcal{A}}\Psi_{\mathsf{A}}^{\left(\upsilon\right)}+i\upsilon g_{\Sigma}\Psi_{\Sigma}^{\left(\upsilon\right)}=0\,.\label{Comoving_Perturbation_eqs:EoS_2_coeff_simpler}
\end{equation}

\subsection{The complete perturbation system\label{subsec:Complete-perturbation-system}}

Equation~\eqref{Comoving_Perturbation_eqs:EoS_2_differential_general_full}
follows from the requirement of having to specify the source of the
anisotropic stresses in the interior of the perturbed stellar object.
As such, the coefficient $g_{\Pi}$ cannot vanish. Therefore, Eq.~\eqref{Comoving_Perturbation_eqs:EoS_2_coeff_simpler}
can be used to replace the coefficient $\Psi_{\mathcal{P}}^{\left(\upsilon\right)}$
in the perturbation equation. Moreover, considering Eq.~\eqref{eqUps}, the constraint equation~\eqref{Comoving_Perturbation_eqs:Extra_constraint_after_EOS} can be used to replace one of the variables $\Psi_{\mathsf{A}}^{\left(\upsilon\right)}$,
$\Psi_{\mathsf{p}}^{\left(\upsilon\right)}$ and $\Psi_{\Sigma}^{\left(\upsilon\right)}$, such that the perturbations are fully characterized by two master variables. Choosing, for instance,
$\Psi_{\mathsf{p}}^{\left(\upsilon\right)}$ and $\Psi_{\Sigma}^{\left(\upsilon\right)}$ as the master variables, we find the general differential equations for the perturbations
\begin{equation}
	\begin{aligned}\widehat{\Psi}_{\mathsf{p}}^{\left(\upsilon\right)} & =\frac{i\upsilon g_{\Sigma}}{\mathcal{H}\left(g_{\Pi}-g_{p}\right)} \left(G_{1}\Psi_{\Sigma}^{\left(\upsilon\right)}+G_2\Psi_{\mathsf{p}}^{\left(\upsilon\right)}\right)+\frac{i\upsilon g_{\Sigma}}{\mathcal{H}\left(g_{\Pi}-g_p\right)} G_3\left(\mathcal{M}\Psi_{\mathsf{p}}^{\left(\upsilon\right)}-\mathcal{N}\Psi_{\Sigma}^{\left(\upsilon\right)}\right)\\
		& +\frac{g_{\Pi}}{g_{\Pi}-g_p}\left[1+\frac{2g_{\Sigma}}{3\left(g_{\Pi}-g_p\right)}\left(\frac{f_p}{f_{\mu}}-\frac{f_{\Pi}g_{p}}{f_{\mu}g_{\Pi}}\right)\frac{i\upsilon}{\mathcal{H}}\right]\mathcal{F}_3\left(\mathcal{M}\Psi_{\mathsf{p}}^{\left(\upsilon\right)}-\mathcal{N}\Psi_{\Sigma}^{\left(\upsilon\right)}\right)\\
		& +\frac{g_{\Pi}}{g_{\Pi}-g_p}\left[1+\frac{2g_{\Sigma}}{3\left(g_{\Pi}-g_{p}\right)}\left(\frac{f_{p}}{f_{\mu}}-\frac{f_{\Pi}g_p}{f_{\mu}g_{\Pi}}\right)\frac{i\upsilon}{\mathcal{H}}\right]\left[\mathcal{F}_1\Psi_{\Sigma}^{\left(\upsilon\right)}+\mathcal{F}_2\Psi_{\mathsf{p}}^{\left(\upsilon\right)}\right]\,,
	\end{aligned}
	\label{Comoving_Perturbation_eqs:Final_system_p_hat}
\end{equation}
\begin{equation}
	\begin{aligned}\mathcal{H}\widehat{\Psi}_{\Sigma}^{\left(\upsilon\right)} & =\frac{2}{3}\left(\frac{f_{p}}{f_{\mu}}-\frac{f_{\Pi}g_{p}}{f_{\mu}g_{\Pi}}\right)\frac{g_{\Pi}}{g_{\Pi}-g_{p}}\left[\mathcal{F}_{1}\Psi_{\Sigma}^{\left(\upsilon\right)}+\mathcal{F}_{2}\Psi_{\mathsf{p}}^{\left(\upsilon\right)}\right]+G_{1}\Psi_{\Sigma}^{\left(\upsilon\right)}+G_{2}\Psi_{\mathsf{p}}^{\left(\upsilon\right)}\\
		& +\left[\frac{2}{3}\left(\frac{f_{p}}{f_{\mu}}-\frac{f_{\Pi}g_{p}}{f_{\mu}g_{\Pi}}\right)\frac{g_{\Pi}}{g_{\Pi}-g_{p}}\mathcal{F}_{3}+G_{3}\right]\left(\mathcal{M}\Psi_{\mathsf{p}}^{\left(\upsilon\right)}-\mathcal{N}\Psi_{\Sigma}^{\left(\upsilon\right)}\right)\,,
	\end{aligned}
	\label{Comoving_Perturbation_eqs:Final_system_Sigma_hat}
\end{equation}
together with Eq.~\eqref{eqUps}, where the functions $\mathcal{F}_{1-3}$, $G_{1-3}$, $\mathcal{M}$, $\mathcal{N}$, and $\mathcal{H}$
are defined in Appendix~\ref{Appendix_sec:Auxiliary_functions_comoving_frame}.
The constraints equations relating the harmonic coefficients of the remaining 1+1+2
potentials with $\Psi_{\mathsf{p}}^{\left(\upsilon\right)}$
and $\Psi_{\Sigma}^{\left(\upsilon\right)}$ read
\begin{equation}
\begin{aligned}\Psi_{\mathsf{E}}^{\left(\upsilon\right)} & =\left[\frac{g_{p}}{2g_{\Pi}}+\frac{\Pi_{0}+6\mathcal{E}_{0}}{6\left(\mu_{0}+p_{0}\right)}\left(\frac{f_{\Pi}g_{p}}{f_{\mu}g_{\Pi}}-\frac{f_{p}}{f_{\mu}}\right)\right]\Psi_{\mathsf{p}}^{\left(\upsilon\right)}+\left[\frac{1}{2}+\frac{\left(\Pi_{0}+6\mathcal{E}_{0}\right)f_{\Pi}}{6\left(\mu_{0}+p_{0}\right)f_{\mu}}\right]\frac{g_{\mathcal{A}}}{g_{\Pi}}\Psi_{\mathsf{A}}^{\left(\upsilon\right)}\\
 & +\left[\left(\frac{1}{2}+\frac{\left(\Pi_{0}+6\mathcal{E}_{0}\right)f_{\Pi}}{6\left(\mu_{0}+p_{0}\right)f_{\mu}}\right)\frac{g_{\Sigma}}{g_{\Pi}}i\upsilon-\frac{1}{2}\left(\mu_{0}+p_{0}-\frac{1}{2}\Pi_{0}-3\mathcal{E}_{0}\right)\left(1+\frac{\Pi_{0}}{\mu_{0}+p_{0}}\right)\right]\Psi_{\Sigma}^{\left(\upsilon\right)}\,,
\end{aligned}
\end{equation}
\begin{equation}
\left(\mu_{0}+p_{0}\right)\Psi_{\theta}^{\left(\upsilon\right)}=-\left(\frac{3}{2}\Pi_{0}+i\upsilon\frac{f_{\Pi}g_{\Sigma}}{f_{\mu}g_{\Pi}}\right)\Psi_{\Sigma}^{\left(\upsilon\right)}+\left(\frac{f_{p}}{f_{\mu}}-\frac{f_{\Pi}g_{p}}{f_{\mu}g_{\Pi}}\right)\Psi_{\mathsf{p}}^{\left(\upsilon\right)}-\frac{f_{\Pi}g_{\mathcal{A}}}{f_{\mu}g_{\Pi}}\Psi_{\mathsf{A}}^{\left(\upsilon\right)}\,,
\end{equation}
\begin{equation}
\begin{aligned}\Psi_{\mathsf{F}}^{\left(\upsilon\right)} & =\frac{\left(2\mathcal{A}_{0}-\phi_{0}\right)}{3\left(\mu_{0}+p_{0}\right)}\left[\frac{f_{p}}{f_{\mu}}-\frac{f_{\Pi}g_{p}}{f_{\mu}g_{\Pi}}\right]\Psi_{\mathsf{p}}^{\left(\upsilon\right)}-\frac{\left(2\mathcal{A}_{0}-\phi_{0}\right)f_{\Pi}g_{\mathcal{A}}}{3\left(\mu_{0}+p_{0}\right)f_{\mu}g_{\Pi}}\Psi_{\mathsf{A}}^{\left(\upsilon\right)}\\
 & -\left[\left(\mathcal{A}_{0}-\frac{1}{2}\phi_{0}\right)\left(1+\frac{\Pi_{0}}{\mu_{0}+p_{0}}\right)+\frac{\left(2\mathcal{A}_{0}-\phi_{0}\right)f_{\Pi}g_{\Sigma}}{3\left(\mu_{0}+p_{0}\right)f_{\mu}g_{\Pi}}i\upsilon\right]\Psi_{\Sigma}^{\left(\upsilon\right)}\,,
\end{aligned}
\end{equation}
\begin{equation}
\Psi_{\mathcal{P}}^{\left(\upsilon\right)}=-\frac{g_{p}}{g_{\Pi}}\Psi_{\mathsf{p}}^{\left(\upsilon\right)}-\frac{g_{\mathcal{A}}}{g_{\Pi}}\Psi_{\mathsf{A}}^{\left(\upsilon\right)}-i\upsilon\frac{g_{\Sigma}}{g_{\Pi}}\Psi_{\Sigma}^{\left(\upsilon\right)}\,,\label{Comoving_Perturbation_eqs:Final_system_P_constraint}
\end{equation}
\begin{equation}
\Psi_{\mathsf{m}}^{\left(\upsilon\right)}=-\frac{f_{p}}{	f_{\mu}}\Psi_{\mathsf{p}}^{\left(\upsilon\right)}-\frac{f_{\Pi}}{	f_{\mu}}\Psi_{\mathcal{P}}^{\left(\upsilon\right)}\,,
\end{equation}
\begin{equation}
	\Psi_{\mathsf{A}}^{\left(\upsilon\right)}=\mathcal{M}\Psi_{\mathsf{p}}^{\left(\upsilon\right)}-\mathcal{N}\Psi_{\Sigma}^{\left(\upsilon\right)}\,.
	\label{Comoving_Perturbation_eqs:Final_system_constraint_A}
\end{equation}

In the derivation of the last form for the perturbation equations,
we have included denominators with $g_{\Pi}-g_{p}$. Therefore, these
equations cannot be used to study the case when $\Pi=p+\text{const}$.
Nonetheless, in that specific case, we can simply use the system~\eqref{Comoving_Perturbation_eqs:A_hat_after_EOS}--\eqref{Comoving_Perturbation_eqs:Mu_after_EOS}
considering $\Psi_{\mathsf{p}}^{\left(\upsilon\right)}=\Psi_{\mathcal{P}}^{\left(\upsilon\right)}$.

Even without solving the system, we can immediately draw some conclusions
on the behavior of the perturbations depending on the specific nature
of the anisotropic ansatz.
If the functions  $f$ and $g$  
do not explicitly depend on the eigenfrequencies, other than the indirect dependency following from dot-derivatives of the kinematic and matter variables, the terms with odd powers of $\upsilon$, in Eqs.\,\eqref{Comoving_Perturbation_eqs:Final_system_p_hat}--\eqref{Comoving_Perturbation_eqs:Final_system_constraint_A}, are all multiplied by the imaginary unit, e.g., by terms like $i\upsilon$, $i\upsilon^{3}$, etc.
Therefore, in general, the Fourier coefficients $\Psi_{\chi}^{\left(\upsilon\right)}$, where $\chi$  is any of the thermodynamic first-order variables, will have real
and imaginary parts. However, from the equations, they verify
\begin{equation}
\begin{aligned}\text{Re}\left(\Psi_{\chi}^{\left(\upsilon\right)}\right) & =\text{Re}\left(\Psi_{\chi}^{\left(-\upsilon\right)}\right)\,,\\
\text{Im}\left(\Psi_{\chi}^{\left(\upsilon\right)}\right) & =-\text{Im}\left(\Psi_{\chi}^{\left(-\upsilon\right)}\right)\,.
\end{aligned}
\end{equation}
When computing the inverse Fourier transform, the imaginary terms' contribution is identically zero: the eigenfunctions of the linear perturbations are always even functions of the radial coordinate and the eigenfrequencies.

Lastly, to check the consistency of the system~\eqref{Comoving_Perturbation_eqs:Final_system_p_hat}--\eqref{Comoving_Perturbation_eqs:Final_system_constraint_A},
we can see that it generalizes the system of perturbation equations
for radial, adiabatic perfect fluids, found in Ref.~\citep{Luz_Carloni_2024b}.
Setting in the background $\Pi_{0}=0$, considering $f\left(\mu,p\right)=0$,
such that $f_{\Pi}=0$, and imposing the anisotropic ansatz $\Pi=0$,
that is $g_{\Pi}=1$, and setting all remaining $g$-coefficients to
zero, yields the radial, adiabatic, isotropic perturbation equations
in~\citep{Luz_Carloni_2024b}.

\subsection{Boundary condition}

Provided the system of differential equations, boundary conditions
are required to select the physically meaningful solutions. In that
regard, we will assume the interior solution is smoothly matched with
an exterior vacuum solution. Then, an observer comoving with the fluid
defines the boundary of the perturbed stellar object as the 3-hypersurface
at which $p+\Pi=0$ at all times, implying $\dot{p}+\dot{\Pi}=0$.
At linear level, applying a harmonic decomposition considering the
eigenfunctions of the Laplace-Beltrami operator and the time derivative
operator of the background spacetime yields the boundary condition
for the harmonic coefficients
\begin{equation}
\Psi_{\mathsf{p}}^{\left(\upsilon\right)}+\Psi_{\mathcal{P}}^{\left(\upsilon\right)}=0\,.\label{Comoving_Perturbation_eqs:Boundary_condition}
\end{equation}
Since we have found that only the real parts of the harmonic coefficients
contribute to the actual solution, the above boundary condition in
fact only has to be ensured on the real part.

\subsection{Breaking covariance\label{subsec:Breaking_covariance}}

Equations~\eqref{Comoving_Perturbation_eqs:Final_system_p_hat}--\eqref{Comoving_Perturbation_eqs:Final_system_constraint_A}
are written covariantly. However, to find particular solutions, it is useful to break covariance and write the system of evolution equations
in a particular coordinate system. At this point, the system is fully
identification gauge-invariant, and such breaking of covariance does not
lead to fictitious solutions. Considering the Schwarzschild coordinate
system defined by a radially static observer at spatial infinity,
with time coordinate $t$ and circumferential radius coordinate $r$,
we can write the differential system~\eqref{Comoving_Perturbation_eqs:Final_system_p_hat} and \eqref{Comoving_Perturbation_eqs:Final_system_Sigma_hat} in matrix form
\begin{equation}
\frac{d\mathds{W}}{dr}=\left(\frac{1}{r}\mathds{R}+\Theta\right)\mathds{W}\,,\label{Comoving_Perturbation_eqs:Matrix_form_general}
\end{equation}
where
\begin{equation}
	\begin{aligned}
		\mathds{W} & =\left[\begin{array}{c}
			\Psi_{\mathsf{p}}^{\left(\upsilon\right)}\\
			\Psi_{\Sigma}^{\left(\upsilon\right)}
		\end{array}\right]\,, & \mathds{R} & =\left[\begin{array}{cc}
			\mathds{R}_{11} & \mathds{R}_{12}\\
			\mathds{R}_{21} & \mathds{R}_{22}
		\end{array}\right]\,, & \Theta & =\left[\begin{array}{cc}
			\Theta_{11} & \Theta_{12}\\
			\Theta_{21} & \Theta_{22}
		\end{array}\right]\,,
	\end{aligned}
	\label{Comoving_Perturbation_eqs:R_matrix_general}
\end{equation}
and imposing that the eigenfrequencies $\upsilon$ are given by Eq.~\eqref{eq:Ups_radial}.
Due to their length, the expressions for the entries of the matrices $\mathds{R}$ and $\Theta$ can be found in Appendix~\ref{Appendix_sec:Auxiliary_functions_comoving_frame}.
The perturbations are completely characterized by the matrix $\Theta$
and the matrix $\mathds{R}$, and the boundary conditions. The matrix
$\mathds{R}$ fully characterizes the singular part of the perturbation
equations, and understanding its properties is crucial for the characterization of
the solutions. It is clear, and in contrast with the isotropic case, that 
the elements of the matrix $\mathds{R}$, Eqs.~\eqref{Appendix_eq:R11}--\eqref{Appendix_eq:R22}, markedly change
depending on the equation of state and the anisotropic ansatz.
Consequently, the nature of the solutions is strongly connected to the properties of the matrix $\mathds{R}$. In the following, 
we will explore the role of anisotropies in linear perturbation theory, considering various ansatze for
$g$ and $f$ and particular background spacetimes.

\section{Anisotropic ansatze from Eckart, BDNK, and Truncated Israel-Stewart theories}\label{Sec:Anisotropic_ansatze}

In general, we require that solutions of the TOV equations for static, spherically symmetric spacetimes are
of class $\mathcal{C}^{2}$. However, most closed form solutions for
stellar compact objects present in the literature are real analytic,
at least in some neighborhood around the center of the star. Indeed,
assuming that the functions $f$ and $g$ are real analytic, and $g_{\Pi}\neq g_{p}$
around the center, we can find the general form of the solutions.
Nonetheless, the nature of the solutions dramatically changes if the
matrix $\mathds{R}$ is or not diagonalizable, or if it has or not
eigenvalues that differ by a non-zero integer. Moreover,
if the eigenvalues differ by a non-zero integer, the relative values
of the eigenvalues is also important to find the general form of the
solution. Therefore, the cases where the matrix $\mathds{R}$ is not
diagonalizable, or has two eigenvalues that differ by a non-zero
integer can only be treated by specifying the anisotropic ansatz and
the equation of state.

In this section, we will consider the particular cases of the anisotropic
ansatz to be such that $g_{\Pi}$ and $g_{\Sigma}$ are not trivial, and all remaining
$g$-coefficients vanish identically. Indeed, the introduction of
the $f$ and $g$ functions provides a way to study anisotropic ansatze
following from very distinct theories in a unified framework. Particular
examples of such ansatze are those following from Eckart theory, the
effective relativistic fluid model from Bemfica-Disconzi-Noronha-Kovtun
(BDNK), and the so-called Truncated Israel-Stewart theory. For further
details, see Appendix~\ref{Appendix_sec:EoS2}. In the Eckart and
the BDNK theories, Eq.~\eqref{Comoving_Perturbation_eqs:EoS_2_coeff}
simplifies to
\begin{equation}
\Psi_{\mathcal{P}}^{\left(\upsilon\right)}+2i\upsilon\eta\Psi_{\Sigma}^{\left(\upsilon\right)}=0\,,\label{Eckart_eq:Ansatz_algebric}
\end{equation}
that is, the only non-trivial $g$-coefficients are $g_{\Pi}=1$ and
$g_{\Sigma}=2\eta$, where $\eta$ is the shear viscosity of the fluid.
In the case of the Truncated Israel-Stewart theory for non-equilibrium
thermodynamics, Eq.~\eqref{Comoving_Perturbation_eqs:EoS_2_coeff}
reads
\begin{equation}
\left(1+i\upsilon\tau_{2}\right)\Psi_{\mathcal{P}}^{\left(\upsilon\right)}+2i\upsilon\eta\Psi_{\Sigma}^{\left(\upsilon\right)}=0\,,\label{Eckart_eq:Ansatz_Israel_Stewart}
\end{equation}
such that the only non-trivial $g$-coefficients are $g_{\Pi}=1+i\upsilon\tau_{2}$
and $g_{\Sigma}=2\eta$, where $\tau_{2}$ is a relaxational time,
defining a characteristic timescale for the perturbation response.

Comparing Eqs.~\eqref{Eckart_eq:Ansatz_algebric} and \eqref{Eckart_eq:Ansatz_Israel_Stewart},
it is immediate to conclude that for $\upsilon=0$ the anisotropic
ansatz is the same for all three theories. Therefore, for a particular
background spacetime, the marginally stable case is independent of
which of these theories is considered. Moreover, the Fourier transformed
version of the anisotropic ansatze makes it clear that the effects
of considering the Truncated version of the Israel-Stewart theory
over the Eckart or BDNK theories become significant when $\left|\upsilon\tau_{2}\right|\gg1$.

Continuing, if the only non-trivial $g$-coefficients are $g_{\Pi}$
and $g_{\Sigma}$, the matrix $\mathds{R}$ simplifies to
\begin{equation}
\mathds{R}=\left[\begin{array}{ccc}
0  & \mathds{R}_{12}\\
0  & \mathds{R}_{22}
\end{array}\right]\,,\label{Eckart_eq:R_matrix}
\end{equation}with
\begin{equation}
	\begin{aligned}
		 \mathds{R}_{12} & =\left[-\frac{2}{\mathcal{H}}\left(\frac{\upsilon g_{\Sigma}}{g_{\Pi}}\right)^{2}\left(\frac{f_{\Pi}}{f_{\mu}}\right)\right]_{r=0}\,,\\
		\mathds{R}_{22} & =\left[\frac{2i\upsilon g_{\Sigma}}{\mathcal{H}g_{\Pi}}\left(\frac{f_{p}}{f_{\mu}}\right)-\frac{3\left(\mu_{0}+p_{0}\right)}{\mathcal{H}}\right]_{r=0}\,.
	\end{aligned}
	\label{Eckart_eq:R_matrix_general_entries}
\end{equation}
where the function $\mathcal{H}$  is defined in Eq.~\eqref{Appencix_eq:H_function}, and we have used the fact that a $\mathcal{C}^{1}$ solution of
the TOV equations is characterized by $\Pi_{0}\left(0\right)=0$.

From the expressions for $\mathds{R}_{12}$ and $\mathds{R}_{22}$, we see
that there are three distinct cases of physical interest: the case
where $f_{\Pi}=0$, the case where $f_{\Pi}\neq0$ but $f_{p}-f_{\Pi}=0$,
and the case where $f_{\Pi}\neq0$ and $f_{p}-f_{\Pi}\neq0$. We shall
consider each of these cases separately.

\subsection{The case where $f$ is independent of $\Pi $}\label{subsec:Eckart-like_ansatze_barotropic-like}

A popular choice in the literature when studying perturbations of
stellar compact objects is to consider a barotropic equation of state,
such that $f\left(\mu,p\right)=0$. In fact, it is usually assumed
an even simpler version, an equation of state of the form $p=f\left(\mu\right)$,
such that $f_{p}=1$, and $f_{\mu}=-f'\left(\mu\right)$ represents,
in absolute value, the square of the adiabatic speed of sound, $c_s^2$.

An equation of state of the form $f\left(\mu,p\right)=0$ implies
$f_{\Pi}=0$. In such case, the $\mathds{R}$ matrix greatly simplifies:
$\mathds{R}_{12}$ vanishes identically and $\mathds{R}_{22}=-3$.
Remarkably, the $\mathds{R}$ matrix is constant with eigenvalues
0 and $-3$. If the $\Theta$ matrix, Eq.~\eqref{Comoving_Perturbation_eqs:Matrix_form_general},
is real analytic in a neighborhood around the center, $r=0$, we can
find the solutions as a power series using the Coddington-Levinson
formalism~\citep{Coddington_Levinson_Book}. Applying the formalism,
we find that the lowest order terms of the power series of the general
solution are
\begin{equation}
	\left[\begin{array}{c}
		\Psi_{\mathsf{p}}^{\left(\upsilon\right)}\\
		\Psi_{\Sigma}^{\left(\upsilon\right)}
	\end{array}\right]=\left[\begin{array}{c}
		-\frac{c_{1}}{r^{2}}\Theta_{12}\left(0\right)\\
		\frac{2c_{1}}{r^{3}}
	\end{array}\right]+\text{higher order terms in }r\,,
\end{equation}
where $\Theta_{12} \left( 0 \right)$ represents the $12$-element of the $\Theta$ matrix evaluated at $r=0$, and $c_{1}$ is an integration constant. We see that for $c_{1}\neq0$
the solutions for the perturbations diverge at the center of the stellar
object at all times. Therefore, to select the physically meaningful
solutions, we must impose that $c_{1}$ vanishes. In that case, we
find that the general solutions around the center, at $r=0$, are of the form
\begin{equation}
	\left[\begin{array}{c}
		\Psi_{\mathsf{p}}^{\left(\upsilon\right)}\\
		\Psi_{\Sigma}^{\left(\upsilon\right)}
	\end{array}\right]=\left[\begin{array}{c}
		c_{2}\\
		\frac{1}{5}c_{2}r^{2}\,\Theta'_{21}\left(0\right)
	\end{array}\right]+\text{higher order terms in }r\,,
	\label{Eckart_eq:Solution_center_barotropic-like}
\end{equation}
where $\Theta'_{21} \left( 0 \right)$ represents the radial derivative of the $21$-element of the $\Theta$ matrix evaluated at $r=0$, and $c_{2}$ is an integration constant. The constant
$c_{2}$ sets the initial amplitude of the perturbation of $\mathsf{p}$
at $r=0$. The value of $c_{2}$
does not affect the eigenfrequencies of the system and simply changes
the overall scale of the amplitudes of the perturbations of the various
quantities that characterize the perturbed stellar compact object.

\subsection{The cases where $f$ depends on $p$  and $\Pi$ in the same way \label{subsec:Eckart_fPi_diff_zero_fp_equal_fPi}}

Another possible choice is to consider an explicit dependency of the
anisotropic pressure in the equation of state of the perturbed fluid,
such that 
\begin{equation}\label{Case2}
f_{\Pi}\neq0 \qquad f_{p}-f_{\Pi}=0\,.
\end{equation}

In the case when $f_{\Pi}\neq0$ and $f_{p}-f_{\Pi}=0$, the $\mathds{R}$
matrix in Eqs.~\eqref{Eckart_eq:R_matrix} and \eqref{Eckart_eq:R_matrix_general_entries}
can be further simplified, such that
\begin{equation}
	\begin{aligned}
			\mathds{R}_{12} & 
			=\left[\frac{-2f_{\Pi}}{\left(\mu_{0}+p_{0}\right)f_{\mu}}\left(\frac{g_{\Sigma}}{g_{\Pi}}\upsilon\right)^{2}\right]_{r=0}\,, &
			&\quad&
			\mathds{R}_{22} & 
			=\left[\frac{2i\upsilon g_{\Sigma}}{\left(\mu_{0}+p_{0}\right)g_{\Pi}}\frac{f_{p}}{f_{\mu}}-3\right]_{r=0}\,.
		\end{aligned}
\end{equation}

Differently from the case of the previous subsection, for $f_{\Pi}\neq0$
the $\mathds{R}$ matrix in Eq.~\eqref{Eckart_eq:R_matrix} will
have complex eigenvalues. To see how this affects the solutions, we
will discuss their behavior around the center. For that, we will assume
that $\mathds{R}_{22}$ is not an integer. 
In this way, the only integer eigenvalue of $\mathds{R}$ is 0. This assumption is physical
reasonable, since $\mathds{R}_{22}$ will be an integer for an extremely
fine-tuned model. Applying the Coddington-Levison approach, around
the center, the general solution will be of the form:
\begin{equation}
	\left[\begin{array}{c}
		\Psi_{\mathsf{p}}^{\left(\upsilon\right)}\\
		\Psi_{\Sigma}^{\left(\upsilon\right)}
	\end{array}\right]=\left[\begin{array}{c}
		c_{2}+c_{1}\frac{\mathds{R_{22}}_{12}}{\mathds{R_{22}}_{22}}\left(r^{\mathds{R_{22}}_{22}}-1\right)\\
		c_{1}r^{\mathds{R_{22}}_{22}}
	\end{array}\right]+\text{higher order terms in }r\,.
	\label{Eckart_eq:Solution_center_general_non_barotropic}
\end{equation}
The terms $r^{\mathds{R}_{22}}$ will diverge at the center. Notice
that even if some eigenfrequencies $\upsilon$ are imaginary, there
will be real positive eigenfrequencies for which the perturbations
will diverge at the center at all times. Therefore, to find the physical
solutions we must set the integration constant $c_{1}$ to zero. In
that case, the physical solutions of the theory in a neighborhood
around the center are given by
\begin{equation}
	\left[\begin{array}{c}
		\Psi_{\mathsf{p}}^{\left(\upsilon\right)}\\
		\Psi_{\Sigma}^{\left(\upsilon\right)}
	\end{array}\right]=\left[\begin{array}{c}
		c_{2}\\
		\frac{c_{2}}{1-\mathds{R_{22}}_{22}}\Theta_{21}\left(0\right)r
	\end{array}\right]+\text{higher order terms in }r\,.
	\label{Eckart_eq:Solution_center_general_non_barotropic_regular}
\end{equation}

From the expressions in Appendix~\ref{Appendix_sec:Auxiliary_functions_comoving_frame}, it is immediate to see that $\Theta_{21} \left( 0 \right)$  is zero if, and only if, $\frac{d}{dr}\left(\frac{f_{p}}{f_{\mu}}\right)=0$. In general, this is not so. However, in the case where the unperturbed fluid is perfect, that is, $\Pi_0$ vanishes identically,  we do have $\Theta_{21}\left(0\right)=0$. That is, in that case, $\Psi_{\Sigma}^{\left(\upsilon\right)} \sim \mathcal{O}\left( r^2\right)$. If the unperturbed fluid is not perfect, $\Theta_{21}\left(0\right)$ might or might not vanish, depending on the equation of state of the perturbed fluid. Nonetheless, it is clear that the presence of anisotropic stresses may markedly change the behavior of the perturbations, even at the center of the stellar object.

\subsection{The cases where $f$ depends on $p$  and $\Pi$ in different ways}

The remaining possibility for the equation of state is that 
\begin{equation}
	\label{Case3}
	\begin{aligned}
		f_{\Pi} & \neq0\,, & \qquad f_{p}-f_{\Pi} & \neq0\,.
	\end{aligned}
\end{equation}
This case is similar to the case of subsection~\ref{subsec:Eckart_fPi_diff_zero_fp_equal_fPi},
where the $\mathds{R}$ matrix in Eqs.~\eqref{Eckart_eq:R_matrix}
and \eqref{Eckart_eq:R_matrix_general_entries} may have complex eigenvalues.
However, the details for choosing the physical solutions are distinct.

Let $\mathds{R}_{22}$ be a non-integer complex number, and consider
an equation of state such that Eq.\,\eqref{Case3} holds.
The solutions in a neighborhood around $r=0$ for the perturbation
equations are still described by Eq.~\eqref{Eckart_eq:Solution_center_general_non_barotropic}.
However, for a general equation of state, from Eq.~\eqref{Eckart_eq:R_matrix_general_entries}, the real part of $\mathds{R}_{22}$ is
\begin{equation}
	\begin{aligned}
			\text{Re}\left(\mathds{R}_{22}\right) & =\left[\frac{4f_{p}\left(f_{\Pi}-f_{p}\right)}{3\mathcal{H}\mathcal{H}^\dag}\left(\frac{g_{\Sigma}}{g_{\Pi}f_{\mu}}\upsilon\right)^{2}-\frac{3\left(\mu_{0}+p_{0}\right)^{2}}{\mathcal{H}\mathcal{H}^\dag}\right]_{r=0}\,,
		\end{aligned}
\end{equation}
where $\mathcal{H}^\dag$ represents the hermitian conjugate of $\mathcal{H}$. Therefore, it is not possible to, in general, infer the sign of the
real part of $\mathds{R}_{22}$. On the one hand, if $\text{Re}\left(\mathds{R}_{22}\right)\leq 0$, the solutions are defined
at $r=0$ if, and only if, $c_{1}=0$, and, in a neighborhood around
the center the lower order terms in $r$ are given by Eq.~\eqref{Eckart_eq:Solution_center_general_non_barotropic_regular}.
On the other hand, the case when $\text{Re}\left(\mathds{R}_{22}\right)> 0$
is more nuanced. 
The Coddington-Levison algorithm only provides the power series solutions around
a singular point, but the series may or may not describe the solutions at
that point. If $c_{1}\neq 0$ the term $r^{\mathds{R}_{22}}$ is not
well-defined at $r=0$, but if $\text{Re}\left(\mathds{R}_{22}\right)>0$
the limit when $r\to0$ exists. Therefore, in that case,
one could choose to consider $c_{1}\neq0$ and extend the solutions
by continuity to $r=0$. Of course, this poses the problem of having
to specify two integration constants.
 We regard this as an interesting mathematical possibility, however
further analysis of such cases will be presented elsewhere.

In the three cases considered in this and the previous subsections,
we have found that the behavior of the physical solutions of the perturbation
equations around the center of the
stellar object is described by Eq.~\eqref{Eckart_eq:Solution_center_barotropic-like} or Eq.~\eqref{Eckart_eq:Solution_center_general_non_barotropic_regular}. As we have discussed, if the background solution is real analytic throughout the interior of the
star, we can use the power series method in~\citep{Coddington_Levinson_Book}.
If this is not possible or is not efficient, the system can be solved
using numerical methods. In that case, Eq.~\eqref{Eckart_eq:Solution_center_barotropic-like}, or Eq.~\eqref{Eckart_eq:Solution_center_general_non_barotropic_regular}, provides the boundary conditions
around the center.

\section{Effects of shear viscosity and relaxational time in the perturbed
eigenmodes}\label{Sec:Effects_shear_relaxation}

Using the new equations for radial adiabatic perturbations, we will
study the effects of shear viscosity and the relaxational time on the
perturbations of stellar compact objects. Concretely, we will analyze
their effects on the eigenfrequencies and the eigenfunctions of the variables $\mathsf{p}$, $\mathsf{A}$, and $\Sigma$, considering
some classical spacetimes widely used in the literature to describe
the interior of stars. Namely, we have considered the Interior Schwarzschild,
the Tolman IV, the Tolman VII~\cite{Tolman_1939}, and the Bowers-Liang (BL) spacetimes~\cite{Bowers_Liang_1974}.
We have opted to include the behavior of $	\Psi_{\mathsf{A}}^{\left(\upsilon\right)}$ to easily compare the results with those found for the isotropic case~\cite{Luz_Carloni_2024b}. In
Table~\ref{Table:Metric_coefficients} we present the non-trivial
metric coefficients in a Schwarzschild coordinate system. We have
considered spacetimes that were also considered in Refs.~\citep{Luz_Carloni_2024b,Luz_Carloni_2024c}
when studying their stability under isotropic, adiabatic perturbations.
This allows us to readily compare the results and analyze the effects
of viscosity and the relaxational time on the eigenmodes. In all examples
we have integrated the perturbation equations using the power series
method and using numerical methods, finding complete agreement to
the considered numerical accuracy.

\begin{table}
\centering
\begin{tabular}{|c|c|}
\hline 
Spacetime & Non-trivial metric components\tabularnewline
\hline 
\hline 
\multirow{1}{*}{Interior Schwarzschild} & $\begin{aligned}\\\left(g_{0}\right)_{tt}= & \left(3\sqrt{1-\frac{2M}{r_{b}}}-\sqrt{1-\frac{2Mr^{2}}{r_{b}^{3}}}\right)^{2}\\
\left(g_{0}\right)_{rr}= & \left(1-\frac{2Mr^{2}}{r_{b}^{3}}\right)^{-1}\\
\\\end{aligned}
$\tabularnewline
\hline 
\multirow{1}{*}{Tolman IV} & $\begin{aligned}\\\left(g_{0}\right)_{tt}= & B^{2}\left(\frac{r^{2}}{A^{2}}+1\right)\\
\left(g_{0}\right)_{rr}= & \frac{\frac{2r^{2}}{A^{2}}+1}{\left(1+\frac{r^{2}}{A^{2}}\right)\left(1-\frac{r^{2}}{R^{2}}\right)}\\
\\\end{aligned}
$\tabularnewline
\hline 
\multirow{1}{*}{Tolman VII} & $\begin{aligned}\\\left(g_{0}\right)_{tt}= & B^{2}\sin^{2}\left[\ln\left(\sqrt{\frac{\sqrt{1-\frac{r^{2}}{R^{2}}+\frac{4r^{4}}{A^{4}}}+\frac{2r^{2}}{A^{2}}-\frac{A^{2}}{4R^{2}}}{C}}\right)\right]\\
\left(g_{0}\right)_{rr}= & \left(1-\frac{r^{2}}{R^{2}}+\frac{4r^{4}}{A^{4}}\right)^{-1}\\
\\\end{aligned}
$\tabularnewline
\hline 
\multirow{1}{*}{Bowers-Liang} & $\begin{aligned}\\\left(g_{0}\right)_{tt}= & A\left[\left(1-\frac{2M}{r_{b}^{3}}r^{2}\right)^{h/2}-3\left(1-\frac{2M}{r_{b}}\right)^{h/2}\right]^{2/h}\\
\left(g_{0}\right)_{rr}= & \left(1-\frac{2M}{r_{b}^{3}}r^{2}\right)^{-1}\\
\\\end{aligned}
$\tabularnewline
\hline 
\end{tabular}\caption{\label{Table:Metric_coefficients}Metric coefficients of classical analytic
solutions of the Einstein field equations. The spacetimes are assumed
to be characterized by a line element of the form~\eqref{Comoving_Perturbation_eqs:general_static_line_element}.}
\end{table}

Following Ref.~\citep{Boyanov_et_al_2024}, we will consider shear
viscosity, $\eta$, and relaxational time, $\tau_{2}$, to be modeled
by
\begin{equation}
\begin{aligned}
	\eta & =\alpha_{\eta}r_{0}\left(p_{0}+ \Pi_{0}\right)\,,
	& \hspace{2em} &
	& \tau_{2} & =\alpha_{\tau}r_{0}\frac{p_{0}+ \Pi_{0}}{\mu_0}\,,
	\end{aligned}
\label{Eckart_eq:Shear_viscosity_ansatz}
\end{equation}
where $p_{0}+ \Pi_{0}$ represents the radial pressure of
the unperturbed fluid (see \eqref{Appendix_ansatze_eq:Radial_tangential_pressures_p_Pi_relations}), and the dimensionless constant $\alpha_{\eta}\in\left]0,\frac{3}{4}\right]$.
Establishing constraints for the value of $\alpha_{\tau}$ is more
involved: the relaxational time depends significantly on the fluid's model
(cf. Ref.~\citep{Boyanov_et_al_2024}). Nonetheless, it is clear
that for $\left|\upsilon\tau_{2}\right|\ll1$, its effects on the
eigenfrequencies and eigenfunctions are negligible compared to those of the Eckart and BDNK models.

In Table\,\ref{Table:Fundamental_eigenfrequecies} we present the values of the fundamental eigenfrequencies for the spacetimes in Table\,\ref{Table:Metric_coefficients},
for different values of $\alpha_{\eta}$ and $\alpha_{\tau}$. As we see, the values of $\alpha_{\eta}$ and $\alpha_{\tau}$ have very little effect on the fundamental eigenfrequencies. These conclusions are in line with those of Ref.\,\cite{Boyanov_et_al_2024}. 

\begin{table}[]
	\centering
	\begin{tabular}{|c|c|c|c||c|c|c||c|c|c||c|c|c|}
		\cline{2-13}
		\multicolumn{1}{c|}{} & \multicolumn{3}{c||}{Int. Schwarzchild} & \multicolumn{3}{c||}{Tolman IV} & \multicolumn{3}{c||}{Tolman VII} & \multicolumn{3}{c|}{Bowers-Liang}\tabularnewline
		\hline 
		{\small\diagbox[width=5em+2\tabcolsep,height=3em]{\quad $\alpha_\tau$\enspace{}}{\quad$\alpha_\eta$ }} & {\small\hspace*{2mm}0.10\hspace*{2mm}} & {\small\hspace*{2mm}0.50\hspace*{2mm}} & {\small\hspace*{2mm}0.75\hspace*{2mm}} & {\small\hspace*{2mm}0.10\hspace*{2mm}} & {\small\hspace*{2mm}0.50\hspace*{2mm}} & {\small\hspace*{2mm}0.75\hspace*{2mm}} & {\small\hspace*{2mm}0.10\hspace*{2mm}} & {\small\hspace*{2mm}0.50\hspace*{2mm}} & {\small\hspace*{2mm}0.75\hspace*{2mm}} & {\small\hspace*{2mm}0.10\hspace*{2mm}} & {\small\hspace*{2mm}0.50\hspace*{2mm}} & {\small\hspace*{2mm}0.75\hspace*{2mm}}\tabularnewline
		\hline 
		\multirow{1}{*}{{\small 0.0}} & {\small 0.108} & {\small 0.108} & {\small 0.108} & {\small 1.533} & {\small 1.534} & {\small 1.536} & {\small 3.434} & {\small 3.436} & {\small 3.439} & {\small 2.072} & {\small 2.079} & {\small 2.087}\tabularnewline
		{\small 2.5} & {\small 0.108} & {\small 0.108} & {\small 0.109} & {\small 1.535} & {\small 1.546} & {\small 1.555} & {\small 3.436} & {\small 3.447} & {\small 3.455} & {\small 2.076} & {\small 2.096} & {\small 2.109}\tabularnewline
		{\small 5.0} & {\small 0.108} & {\small 0.109} & {\small 0.109} & {\small 1.538} & {\small 1.556} & {\small 1.569} & {\small 3.437} & {\small 3.452} & {\small 3.462} & {\small 2.078} & {\small 2.100} & {\small 2.113}\tabularnewline
		\hline 
	\end{tabular}
	
	\caption{\label{Table:Fundamental_eigenfrequecies}Absolute values
		of the fundamental eigenfrequencies, $\left|\lambda_{0}\right|$,
		rounded to three decimal places, assuming the anisotropic ansatze
		following from Eckart and BDNK theories, $\alpha_{\tau}=0$, and the
		truncated Israel-Stewart theory, $\alpha_{\tau}\protect\neq0$. We
		considered an Interior Schwarzschild solution with parameters $\left(M,r_{b}\right)=\left(0.1,1\right)$
		and $c_{s}^{2}=0.1$ for the perturbed fluid, a Tolman IV spacetime
		with parameters $\left(A,B,R\right)=\left(1,1,1.5\right)$, a Tolman
		VII spacetime with parameters $\left(A,B,C,R\right)=\left(1,1,20,0.54\right)$,
		and a Bowers-Liang spacetime with parameters $\left(M,r_{b},h\right)=\left(0.05,1,6\right)$
		and $c_{s}^{2}=0.5$ for the perturbed fluid. For the two Tolman solutions, we have assumed that the background and the perturbed fluid have the same equation of state.}
\end{table}

The low positive correlation between the fundamental eigenfrequencies
and the values of $\alpha_{\eta}$ and of $\alpha_{\tau}$, however,
is not necessarily verified for higher-order eigenmodes. Indeed, we
have found that the effect on the values of the eigenfrequencies is
more significant for higher eigenmodes. Moreover, we found that for
higher eigenmodes, fixing $\alpha_{\eta}$ and increasing $\alpha_{\tau}$
does not always imply an increase in the eigenfrequency.

\begin{figure}
\centering
\includegraphics[width=1\columnwidth]{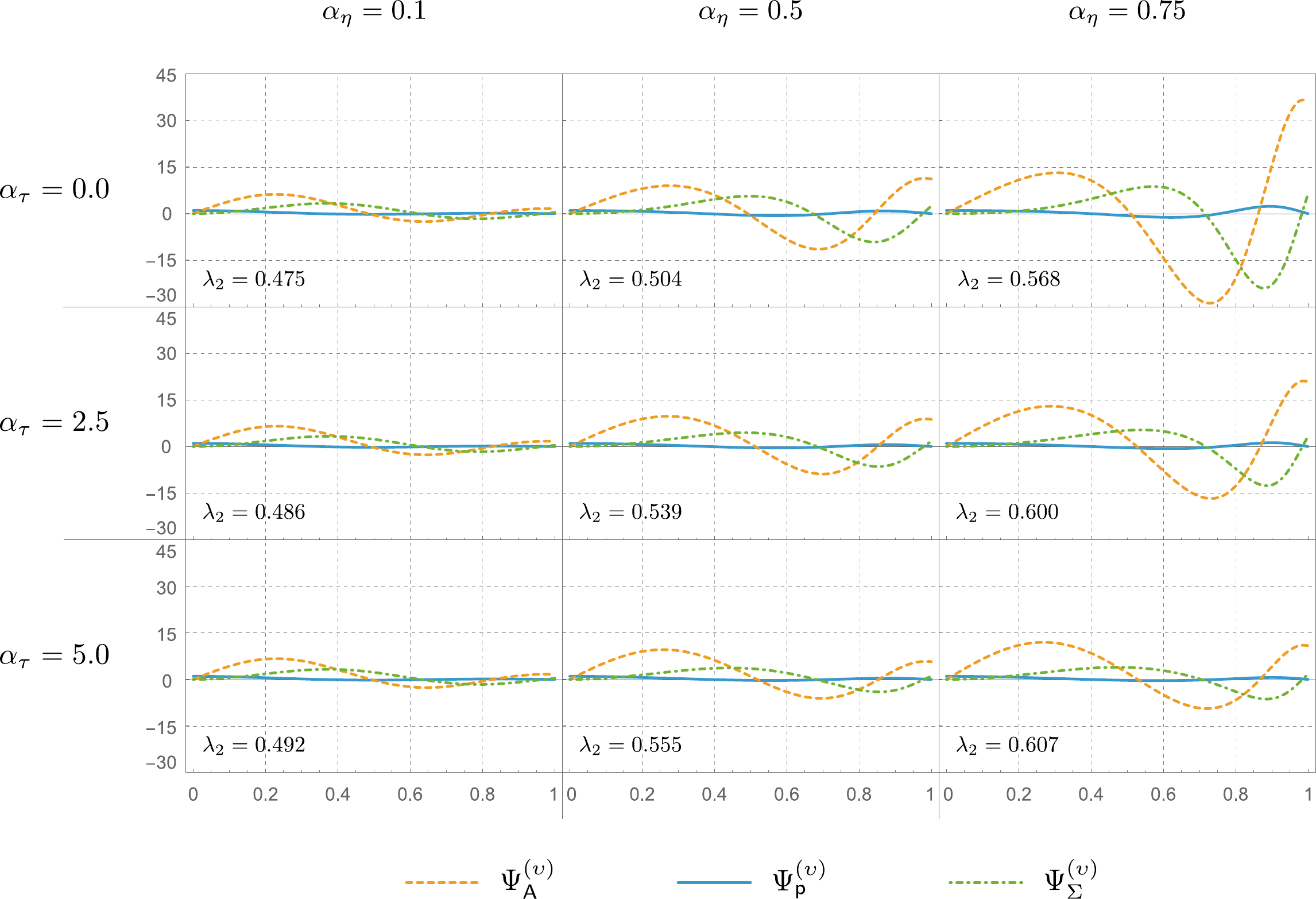}

\caption{\label{Figure:Radial_profile_eigenfunctions_Interior_Schwazschild}Radial
profile of the real Fourier coefficients of the functions $\mathsf{p}$,
$\mathsf{A}$, and $\Sigma$, associated with the third eigenfrequency,
$\lambda_{2}$, for the Interior Schwarzschild spacetime for various
values of the shear viscosity and the relaxational time parameters.
For all cases, the Interior Schwarzschild spacetime parameters are
$\left(M,r_{b}\right)=\left(0.1,1\right)$, and we have set the integration
constant $c_{2}=1$. The speed of sound of the perturbed spacetime
was considered to be $c_{s}^{2}=0.1$. In all graphs, the horizontal
axis represents the values of $r/r_{b}$, where $r_b$ represents the circumferential radius of the unperturbed star.}
\end{figure}

\begin{figure}
\centering
\includegraphics[width=1\columnwidth]{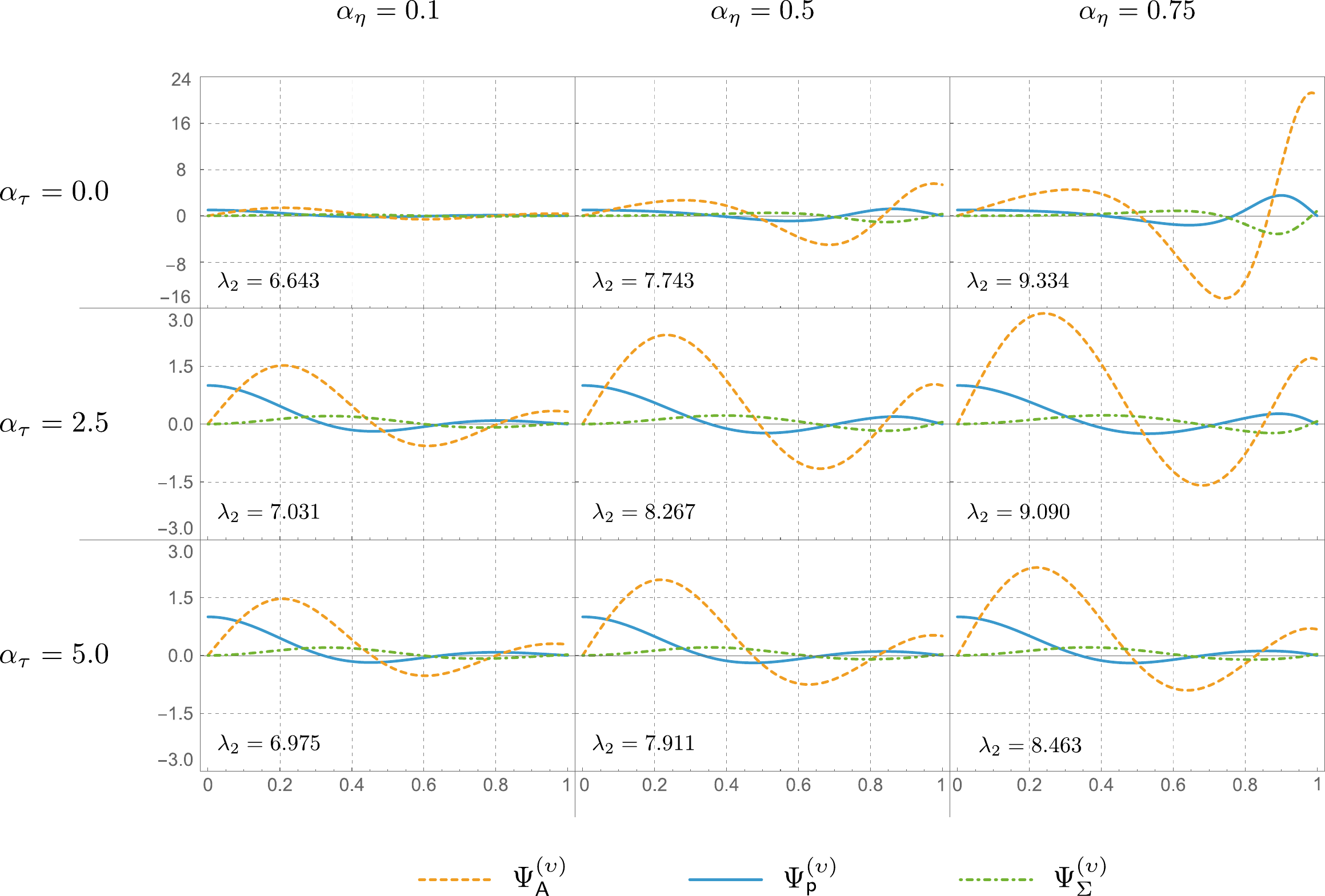}\caption{\label{Figure:Radial_profile_eigenfunctions_Tolman_IV}Radial profile
of the real Fourier coefficients of the functions $\mathsf{p}$, $\mathsf{A}$,
and $\Sigma$, associated with the third eigenfrequencies, $\lambda_{2}$,
for the Tolman IV spacetime for various values of the shear viscosity
and the relaxational time parameters. For all cases, the Tolman IV
spacetime parameters are $\left(A,B,R\right)=\left(1,1,1.5\right)$,
and we have set the integration constant $c_{2}=1$. The perturbed
and unperturbed fluids were considered to have the same equation of
state. In all graphs, the horizontal axis represents the values of
$r/r_{b}$, where $r_b$ represents the circumferential radius of the unperturbed star. We remark that the different range of the vertical axis in
the graphs of the first row.}
\end{figure}

\begin{figure}
\centering
\includegraphics[width=1\columnwidth]{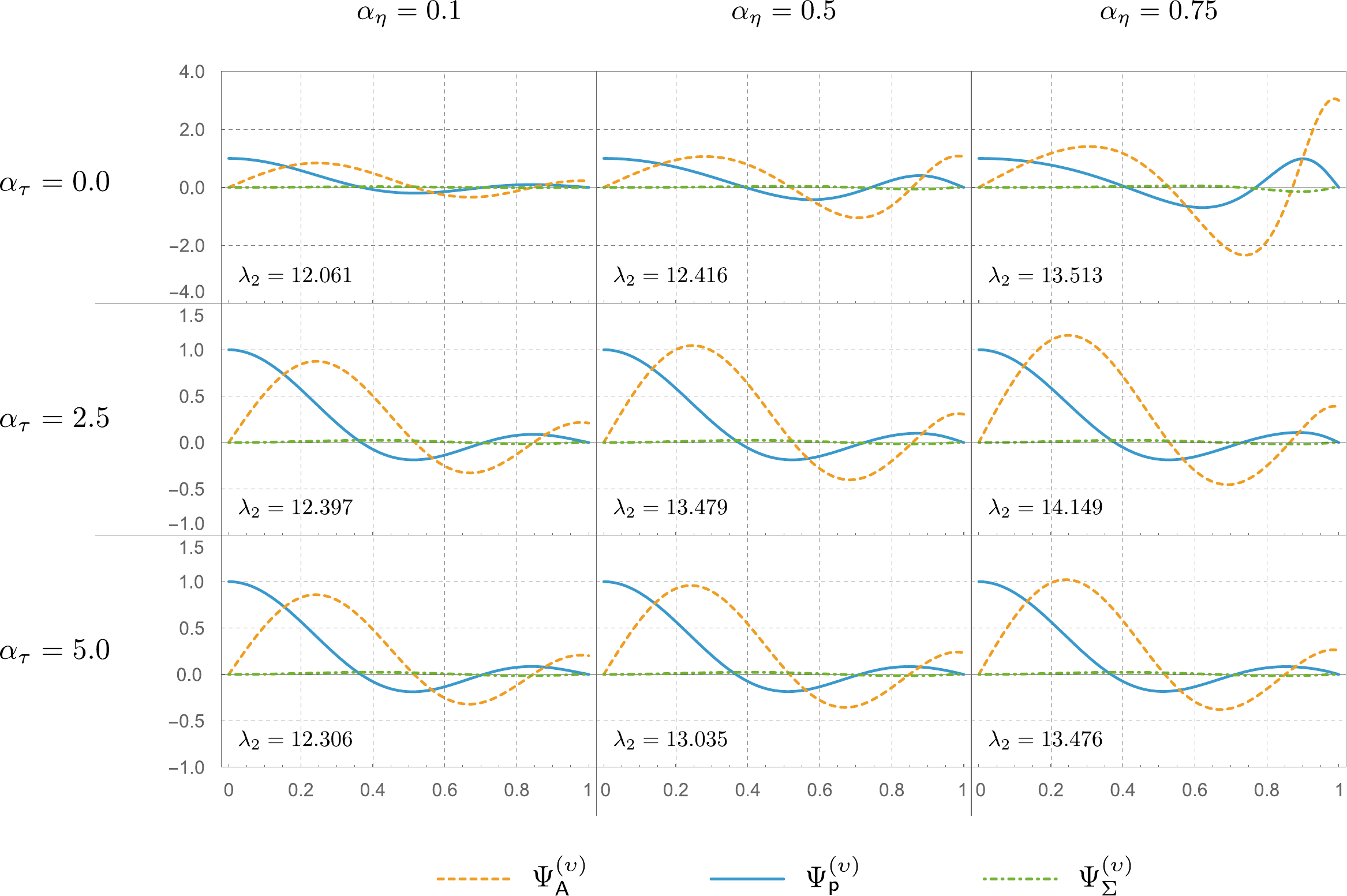}

\caption{\label{Figure:Radial_profile_eigenfunctions_Tolman_VII}Radial profile
of the real Fourier coefficients of the functions $\mathsf{p}$, $\mathsf{A}$,
and $\Sigma$, associated with the third eigenfrequencies, $\lambda_{2}$,
for the Tolman VII spacetime for various values of the shear viscosity
and the relaxational time parameters. For all cases, the Tolman VII
spacetime parameters are $\left(A,B,C,R\right)=\left(1,1,20,0.54\right)$,
and we have set the integration constant $c_{2}=1$. The perturbed
and unperturbed fluids were considered to have the same equation of
state. In all graphs, the horizontal axis represents the values of
$r/r_{b}$, where $r_b$ represents the circumferential radius of the unperturbed star. We remark the different range of the vertical axis in
the graphs of the first row.}

\end{figure}

\begin{figure}
\centering
\includegraphics[width=1\columnwidth]{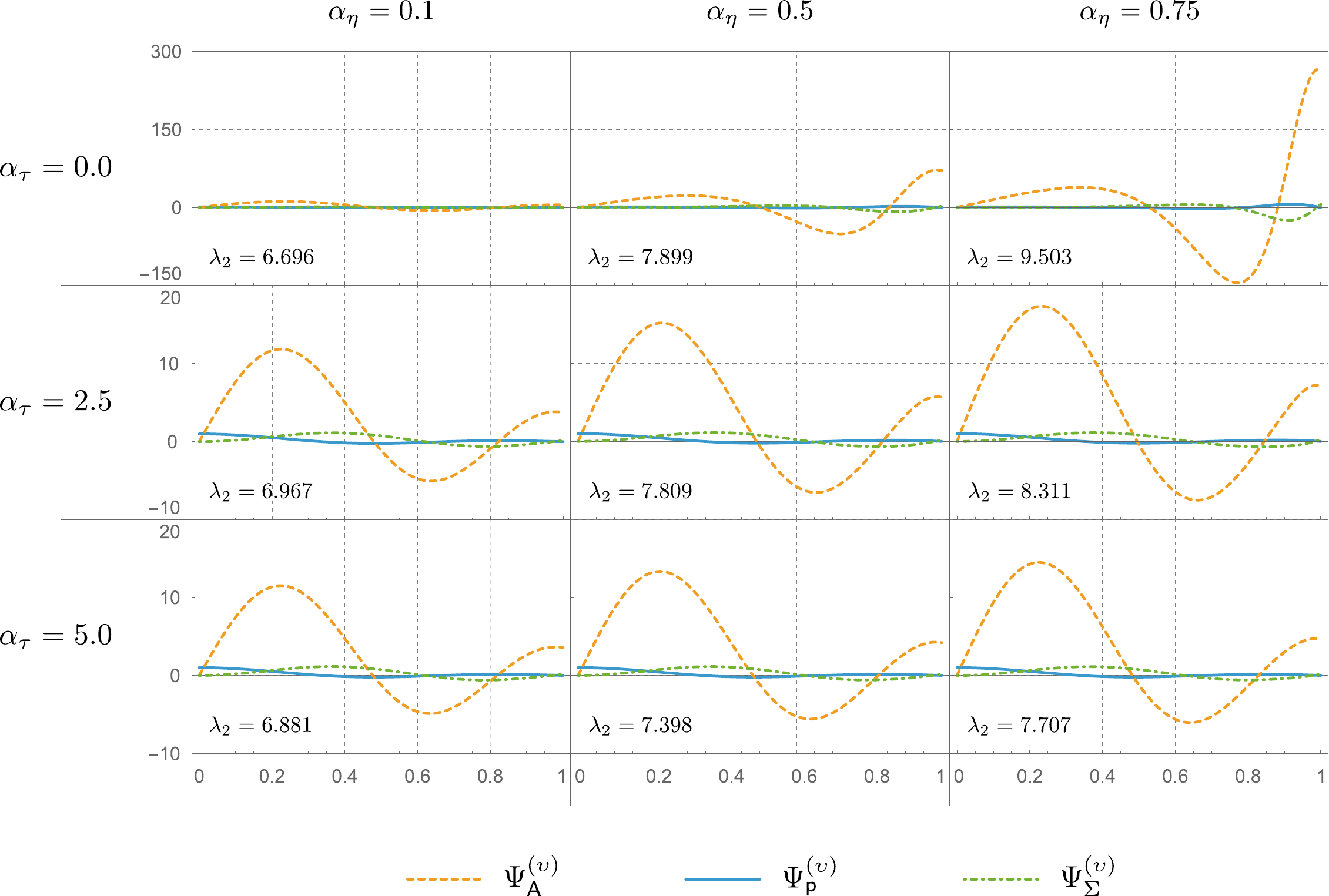}

\caption{\label{Figure:Radial_profile_eigenfunctions_Bowers_Liang}Radial profile
of the real Fourier coefficients of the functions $\mathsf{p}$, $\mathsf{A}$,
and $\Sigma$, associated with the third eigenfrequencies, $\lambda_{2}$,
for the Bowers-Liang spacetime for various values of the shear viscosity
and the relaxational time parameters. For all cases, the Bowers-Liang
spacetime parameters are $\left(M,r_{b},h\right)=\left(0.05,1,6\right)$,
and we have set the integration constant $c_{1}=1$. The speed of
sound of the perturbed spacetime was considered to be $c_{s}^{2}=0.5$.
In all graphs, the horizontal axis represents the values of $r/r_{b}$, where $r_b$ represents the circumferential radius of the unperturbed star. We highlight the different range of the vertical axis in the graphs of
the first row.}
\end{figure}

In Figures~\ref{Figure:Radial_profile_eigenfunctions_Interior_Schwazschild}
-- \ref{Figure:Radial_profile_eigenfunctions_Bowers_Liang} we present
the absolute value of the third eigenfrequency, $\left|\lambda_{2}\right|$,
and its corresponding eigenfunctions of the perturbation variables, for
particular realizations of the spacetimes in Table~\ref{Table:Metric_coefficients} and
considering various values of the constants $\alpha_{\eta}$ and $\alpha_{\tau}$. 
We have chosen to show this eigenmode because of its higher number of nodes,  
which shows the effects on the amplitude of the eigenfunctions more clearly.

In general, for a fixed value of $\alpha_{\tau}$, increasing shear viscosity increases the values of the eigenfrequencies. However, the opposite is not necessarily true. In Figures~\ref{Figure:Radial_profile_eigenfunctions_Tolman_IV}
-- \ref{Figure:Radial_profile_eigenfunctions_Bowers_Liang} we see
that increasing the value of $\alpha_{\tau}$ may yield a greater
or smaller value of $\left|\lambda_{2}\right|$ for a fixed value
of $\alpha_{\eta}$. Moreover, we see that the corrections to the
values of the eigenfrequencies and the eigenfunctions due to shear viscosity and the relaxational
time can be important, compared to the case of $\eta=0=\tau_{2}$.

Considering the anisotropic ansatz following from the Eckart 
and BDNK theories, that is $\tau_{2}=0$ ($\alpha_\tau=0$), in the first row of
Figures~\ref{Figure:Radial_profile_eigenfunctions_Interior_Schwazschild}
-- \ref{Figure:Radial_profile_eigenfunctions_Bowers_Liang}, we see that in the absence
of a relaxation mechanism, and increasingly higher shear viscosity,
the amplitude of the perturbations may grow quite significantly toward
the boundary of the star, even considering a relatively small initial
perturbation at the center.
However,  the magnification of
the amplitude of the eigenfunctions at the boundary is present for
all computed eigenmodes and becomes increasingly more prominent for
higher-order eigenmodes. This result implies that, considering the
Eckart or the BDNK fluid models, a perturbation of a stellar compact
object that excites higher-order eigenmodes would result in a perturbation
whose amplitude around the core is many orders of magnitude smaller
than at the boundary.

The graphs in the second and third row of Figures~\ref{Figure:Radial_profile_eigenfunctions_Interior_Schwazschild} -- \ref{Figure:Radial_profile_eigenfunctions_Bowers_Liang} represent eigenfunctions considering
a fluid with $\tau_{2}\neq0$ ($\alpha_\tau\neq0$), that is, considering the anisotropic
ansatz following from the Truncated Israel-Stewart theory. We see
that the inclusion of a relaxation mechanism tends to soften the perturbations
throughout the interior of the star. Indeed, for higher values of
$\tau_{2}$, shear viscosity seems to have a significantly smaller
effect on the eigenfunctions of radial adiabatic perturbations when
considering the Truncated Israel-Stewart theory than when considering
the Eckart and BDNK fluid models.\textbf{}

\section{Strange matter stellar compact objects}\label{Sec:Strange_stars}

An illustrative example of an equation of state of the form~\eqref{Case2}
is the one following from the MIT bag model. The MIT bag model was
introduced to characterize the structure of hadrons, treating the
three quarks as massless, non-interacting particles confined in a region, a
``bag'', of constant energy density, $\mathcal{B}$, the bag constant.
In this model, the hadron is composed of a Fermi gas with energy density
$\mu$, and radial pressure, $p_{r}$. The constant energy density,
$\mathcal{B}$, behaves dynamically as an external, counterbalancing
pressure and maintains the quark gas at finite density and chemical
potential, such that the fluid is characterized by an equation of
state of the form
\begin{equation}
	\mu=3p_{r}+4\mathcal{B}\,.
\end{equation}
Using Eq.~\eqref{Appendix_ansatze_eq:Radial_tangential_pressures_p_Pi_relations}, this equation of state can be rewritten
as $3p+3\Pi+4\beta-\mu=0$. In terms of the $f$ function, Eq.~\eqref{Comoving_Perturbation_eqs:EoS_general},
the equation of state following from the MIT bag model is characterized
by the coefficients $f_{\mu}=-1$, $f_{p}=3$, and $f_{\Pi}=3$. 

In Ref.~\cite{Witten_1984}, it was conjectured that strange quark
matter, that is, quark matter with strangeness per baryon of order
unity, is the true ground-state of strongly interacting matter, rather
than nuclear matter. In Ref.~\cite{Farhi_Jaffe_1984}, under reasonable
assumptions, it was shown that three-flavor, $\left(u,d,s\right)$,
quark matter is stable for values of the bag constant, $\mathcal{B}$,
between $57\text{ Mev/fm}^{3}$ and $94\text{ Mev/fm}^{3}$. A consequence
of the conjecture and those results is the possible existence of stellar
compact objects composed of strange quark matter, dubbed Strange Stars.

Previous studies have analyzed the properties of Strange Stars, in
particular, their dynamical stability under perturbations, assuming
multiple anisotropic ansatze for the fluid source~\cite{Vath_Chanmugam_1992,Gondek_Zdunik_1999,Arbanil_Malheiro_2016,Mondal_Bagchi_2025}.
However, to our knowledge, there are no results on the properties of the
perturbation of Strange Stars considering the anisotropic ansatze
following from non-equilibrium thermodynamic theories. In this subsection,
we will study the properties of radial, adiabatic, linear perturbations
of Strange Stars, considering the Eckart, BDNK, and the Truncated Israel-Stewart
non-equilibrium thermodynamic theories. Nonetheless, the analysis
presented here is meant to be illustrative of the newly derived integrated
perturbation scheme and not a comprehensive study on the dynamical
stability of Strange Stars. An in-depth study requires analysis of
the properties of the perturbed fluid source, depending on the imposed
anisotropic ansatze for the unperturbed fluid. Instead, we will consider
the simpler setup of an unperturbed static, isotropic Strange Star,
modeled by a solution of the TOV equations for given values of the
central energy density and the bag constant, and study the properties
of the perturbed star depending on the values of shear viscosity,
$\eta$, and relaxational time, $\tau_{2}$.
To allow easy comparison with the results in the literature on
Strange Stars, in this subsection, we will use different units for the various quantities instead of the geometrized unit
system used throughout this article.

\begin{figure}
	\includegraphics[width=1\linewidth]{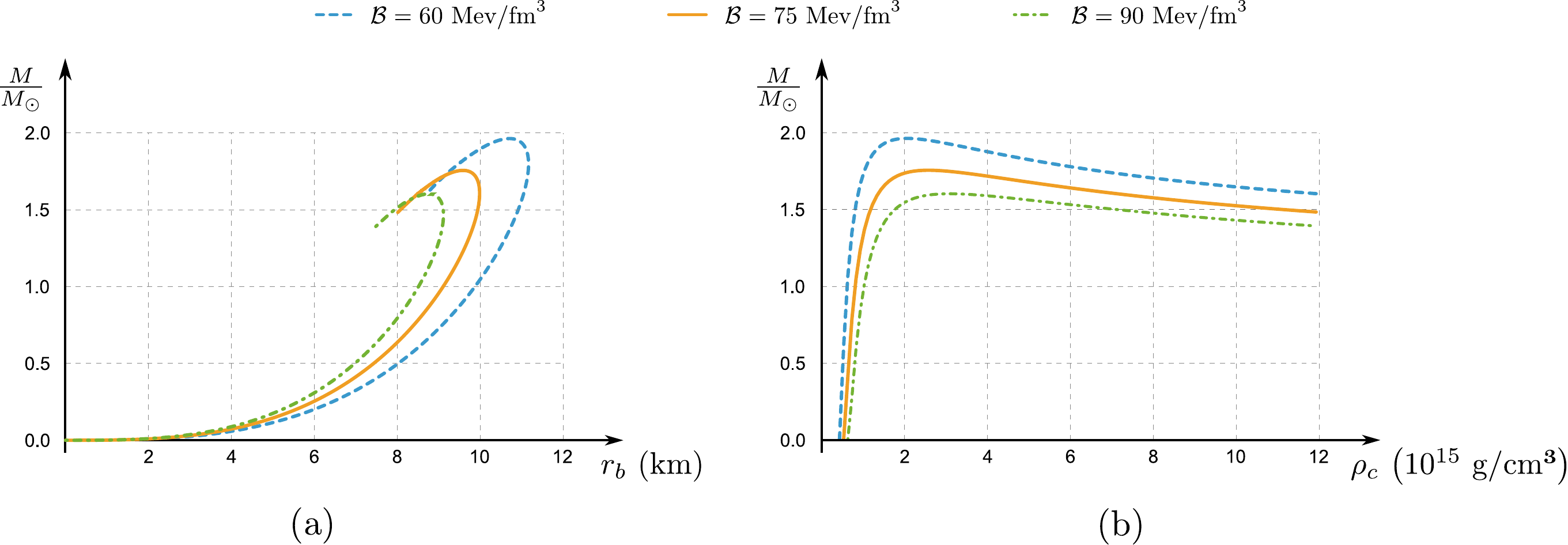}
	
	\caption{\label{Figure:Properties_stellar_object_MIT_bag_model_central_density}Properties
		of a stellar compact object composed of a fluid characterized by an
		equation of state given by the MIT bag model. (a) Mass-radius plot.
		(b) Mass-central density plot.}
	
\end{figure}

In Figure~\ref{Figure:Properties_stellar_object_MIT_bag_model_central_density},
we show the mass-radius profile and the mass as a function of the
central density, characterizing isotropic Strange Stars, for distinct
values of the bag constant.

We consider the linear stability of a perturbed Strange Star for a
fixed value of the central energy density. In Table~\ref{Table:Fundamental_eigenfrequecies_MIT_bag_model}
we list the values of the fundamental eigenfrequencies for various values
of the bag constant, shear viscosity, and the relaxational time, following the model 
in Eq.~\eqref{Eckart_eq:Shear_viscosity_ansatz}. Similarly to
the analytical spacetimes considered in the previous subsection, we
conclude that, for fixed $\mathcal{B}$, the increase of shear viscosity
and relaxational time implies a stabler configuration for a Strange
Star under linear perturbations. However, the change to the fundamental
eigenfrequency remains marginal. Moreover, we see that, for fixed
central density, a strange compact stellar matter is stabler for greater
values of the bag constant. 

\begin{figure}
	\centering\includegraphics[totalheight=0.16\paperheight]{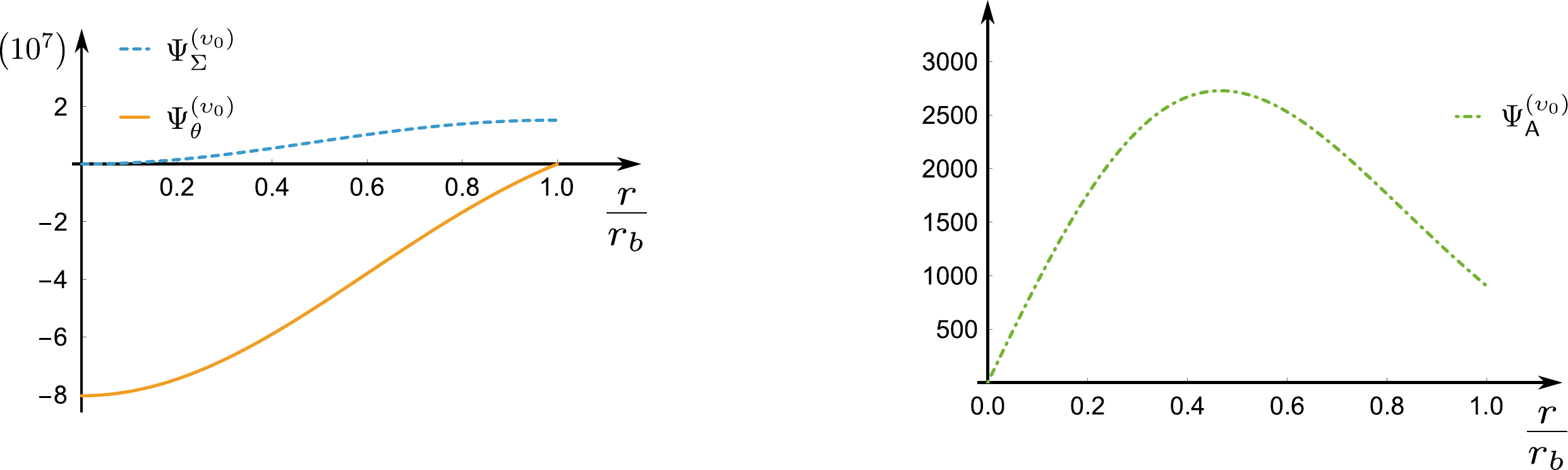}
	
	\caption{\label{Figure:MIT_bag_model_expansion_shear_graphs}Radial profiles
		of the shear and expansion scalars, and the radial component of the
		4-acceleration of a perturbed stellar compact object composed of a
		fluid characterized by an equation of state given by the MIT bag model,
		assuming the bag constant $\mathcal{B}=60\text{ Mev/fm}^{3}$, central
		energy density of the unperturbed fluid $\mu_{c}=900\text{Mev/fm}^{3}$,
		$\alpha_{\eta}=0.5$ and $\alpha_{\tau}=2.5$.}
\end{figure}

An interesting detail concerns the range of applicability of linear perturbation theory. In Figure~\ref{Figure:MIT_bag_model_expansion_shear_graphs} we present
the Fourier coefficients for the fundamental eigenmode of the shear
scalar, $\Psi_{\Sigma}^{\left(\upsilon_{0}\right)}$, the expansion
scalar, $\Psi_{\theta}^{\left(\upsilon_{0}\right)}$, and the time
derivative of the radial component of the 4-acceleration, $\Psi_{\mathsf{A}}^{\left(\upsilon_{0}\right)}$,
that characterize the perturbed spacetime, as functions of the circumferential
radius coordinate. As an example, we show the coefficients for a configuration
with bag constant $\mathcal{B}=60\text{ Mev/fm}^{3}$, central energy
density of the unperturbed fluid $\mu_{c}=900\text{Mev/fm}^{3}$,
$\alpha_{\eta}=0.5$ and $\alpha_{\tau}=2.5$, however, we have found
that the overall behavior and magnitude are similar for other values
of these parameters.

It is clear from these graphs that the values of $\Psi_{\Sigma}^{\left(\upsilon_{0}\right)}$,
$\Psi_{\theta}^{\left(\upsilon_{0}\right)}$ and $\Psi_{\mathsf{A}}^{\left(\upsilon_{0}\right)}$
are not compatible with the assumptions of first-order perturbation
theory: their magnitude is extremely large. These results might be a consequence of imposing the unperturbed
fluid to be perfect, requiring that the perturbations be done adiabatically,
which would imply that a proper linear stability analysis for Strange
Stars must allow the existence of dissipative heat flows, or the necessity
to consider the full non-linear field equations to study the dynamical
stability of Strange Stars. In either case, we see in this example
and those of the previous subsection, that when studying the linear
stability of stellar compact objects, it is pivotal to analyze both
the eigenfrequencies and the eigenfunctions, in the sense that even
if the eigenfrequencies are real, the amplitude of the perturbations
may lie outside the regime of applicability of the linearized theory, making the perturbed analysis irrelevant.

\begin{table}[h]
	{\small{}%
		\begin{tabular}{|c|c|c|c||c|c|c||c|c|c|}
			\cline{2-10}
			\multicolumn{1}{c|}{} & \multicolumn{3}{c||}{{\small$\mathcal{B}=60\text{ Mev/fm}^{3}$}} & \multicolumn{3}{c||}{{\small$\mathcal{B}=75\text{ Mev/fm}^{3}$}} & \multicolumn{3}{c|}{{\small$\mathcal{B}=90\text{ Mev/fm}^{3}$}}\tabularnewline
			\hline 
			{\small\diagbox[width=5em+2\tabcolsep,height=3em]{\quad $\alpha_\tau$\enspace{}}{\quad$\alpha_\eta$ }} & {\small\hspace*{2mm}0.10\hspace*{2mm}} & {\small\hspace*{2mm}0.50\hspace*{2mm}} & {\small\hspace*{2mm}0.75\hspace*{2mm}} & {\small\hspace*{2mm}0.10\hspace*{2mm}} & {\small\hspace*{2mm}0.50\hspace*{2mm}} & {\small\hspace*{2mm}0.75\hspace*{2mm}} & {\small\hspace*{2mm}0.10\hspace*{2mm}} & {\small\hspace*{2mm}0.50\hspace*{2mm}} & {\small\hspace*{2mm}0.75\hspace*{2mm}}\tabularnewline
			\hline 
			\multirow{1}{*}{{\small 0.0}} & 1.329 & 1.356 & 1.391 & 2.184 & 2.218 & 2.262 & 3.031 & 3.068 & 3.116\tabularnewline
			{\small 2.5} & 1.345 & 1.440 & 1.521 & 2.204 & 2.315 & 2.403 & 3.052 & 3.169 & 3.259\tabularnewline
			{\small 5.0} & 1.357 & 1.497 & 1.604 & 2.216 & 2.368 & 2.475 & 3.065 & 3.221 & 3.329\tabularnewline
			\hline 
	\end{tabular}}{\small\par}
	
	\caption{\label{Table:Fundamental_eigenfrequecies_MIT_bag_model}Approximate
		absolute values of the fundamental eigenfrequencies, $\left|\lambda_{0}\right|$,
		in kHz, measured by a static observer at spatial infinity, of a stellar
		compact object composed of a fluid characterized by an equation of
		state given by the MIT bag model, for three distinct values of the
		bag constant, $\mathcal{B}$, and assuming the anisotropic ansatze
		following from Eckart and BDNK theories, $\alpha_{\tau}=0$, and the
		truncated Israel-Stewart theory, $\alpha_{\tau}\protect\neq0$. In
		all cases, we have considered the central energy density $\mu_{c}=900\text{Mev/fm}^{3}$,
		corresponding to central mass density $\rho_{c}\approx1.604\times10^{15}\text{g/cm\ensuremath{^{3}}}$.}
\end{table}

\section{Maximum compactness}\label{Sec:Maximum_compactness}

In Ref.~\citep{Luz_Carloni_2024b}, a general upper bound for the
compactness of perfect fluid stellar compact objects was proposed,
beyond which any object is unstable for isotropic, adiabatic perturbations,
imposing that the perturbed fluid is also perfect. Nonetheless, stellar
compact objects that are composed of non-perfect fluids or self-gravitating
perfect fluids whose radial perturbation generates anisotropic stresses
are expected to be stable for higher compactness values~[15].

To study the maximum compactness of non-perfect fluid stars, we will
follow a similar strategy to the one in Ref.~\citep{Luz_Carloni_2024b}
and consider the marginally stable configuration of solutions of the
Einstein field equations that directly depend on the compactness parameter
$M/r_{b}$, where $M$ is the Schwarzschild gravitational mass parameter
and $r_{b}$ the radius of the star, and have constant energy density
throughout the star. This last requisite is important since it implies
that for a barotropic equation of state, at the linear level, the function
$f_{\mu}$ is simply, in absolute value, the constant adiabatic speed
of sound, allowing us to study the problem without specifying the
equation of state.

Now, there are various solutions of the TOV equations with a non-perfect fluid source with constant energy density.
One such solution is the BL solution~\cite{Bowers_Liang_1974} (see Table~\ref{Table:Metric_coefficients}).
This solution, however, has a non-trivial dependence on the $h$ parameter,
such that, for particular values of $h$, it is unstable
for relatively low values of the compactness parameter. In fact, given the expressions for the radial and tangential pressures of the fluid source of the BL spacetime
\begin{equation}
	\begin{aligned}
		p_r & = -\frac{6M}{r_b^3}
		\frac{\left(1-\frac{2Mr^2}{r_b^3}\right)^{h/2}-\left(1-\frac{2M}{r_b}\right)^{h/2}}{\left(1-\frac{2Mr^2}{r_b^3}\right)^{h/2}-3\left(1-\frac{2M}{r_b}\right)^{h/2}}\,,\\
		p_t-p_r & = \frac{12\left(1-h\right)M^2r^2\left(1-\frac{2M}{r_b}\right)^{h/2}\left(1-\frac{2Mr^2}{r_b^3}\right)^{h/2}}{r_b^6\left(1-\frac{2M}{r_b^3}r^2\right)\left[\left(1-\frac{2Mr^2}{r_b^3}\right)^{h/2}-3\left(1-\frac{2M}{r_b}\right)^{h/2}\right]^2}\,,
	\end{aligned}
\end{equation}
it is straightforward  to show that for a positive energy density, if the parameter $h>0$, and
\begin{equation}
	\frac{2M}{r_{b}} \geq 1-\left(\frac{1}{9}\right)^{\frac{1}{h}}\,,
	\label{Compactness_eq:BL_inequality}
\end{equation}
the BL solution is characterized by divergent radial and tangential pressures at
some $r\in \left[0,r_b\right[$. Naturally, at that radius, the Ricci scalar is
infinite. Therefore, for $h>0$, the value of the compactness parameter in the BL
solution cannot be arbitrarily close to the black hole limit. For $h<1$, the BL spacetime is permeated by a 
fluid with negative anisotropic pressure, and for $h<0$, both the radial and the 
tangential pressures will be negative in a region around the center of the star. We 
remark, however, that in those cases, for a positive energy density, the fluid 
source of the BL solution does not violate the weak nor the strong energy 
conditions.

Given the properties of the BL solution discussed above, we will instead
consider a different solution with constant energy density, derived for the first time in \cite{Carloni:2017bck,Luz:2026fzu}\footnote{We remark, however, that the solution presented in the reference \cite{Carloni:2017bck} has a different (albeit equivalent) form, and it
is more general than the one considered here, as here we have made a specific choice on the parameters.}. Given a line element of the form~\eqref{Comoving_Perturbation_eqs:general_static_line_element}, let
\begin{equation}
	\begin{aligned}
		g_{tt} & =\frac{A\sqrt{1-B\frac{r^{2}}{r_{b}^{2}}}\left(\sqrt{1-B\frac{r^{2}}{r_{b}^{2}}}-3\sqrt{1-B}\right)^{2}}{\sqrt{1-\frac{2M}{r_{b}^{3}}r^{2}}}\,, & g_{rr} & =\left(1-\frac{2Mr^{2}}{r_{b}^{3}}\right)^{-1}\,,
	\end{aligned}
\label{Compactness_eq:Generalized_Interior_Schwarzschild_metric}
\end{equation}
with $A,r_{b}\in\mathbb{R}_{>0}$, $B<\frac{8}{9}$ and $M/r_{b}\in\left[0,\frac{1}{2}\right[$.
For completeness, below we present the expressions for the radial and tangential pressure of the fluid source for this class of solutions:
\begin{equation}
\begin{split}
 p_r & =
 \frac{3B\left[4\sqrt{1-B}\sqrt{1-\frac{Br^{2}}{r_{b}^{2}}}+B\left(\frac{r^{2}}{r_{b}^{2}}+3\right)-4\right]\left(\frac{2Mr^{2}}{r_{b}^{3}}-1\right)}
 {r_{b}^{2}\left[6\sqrt{1-B}\sqrt{1-\frac{Br^{2}}{r_{b}^{2}}}+B\left(\frac{r^{2}}{r_{b}^{2}}+9\right)-10\right]\left(1-\frac{Br^{2}}{r_{b}^{2}}\right)}
 \,,\\
  p_t -p_r & = -\frac{3r^{2}\left(2M-Br_{b}\right)\left(4BMr^{2}-Br_{b}^{3}-2Mr_{b}^{2}\right)}{4\left(r_{b}^{2}-Br^{2}\right)^{2}\left(r_{b}^{4}-2Mr^{2}r_{b}\right)}\,,
\end{split}
\end{equation} 
This solution directly generalizes the Interior Schwarzschild spacetime,
in the sense that setting $B=\frac{2M}{r_{b}}$ implies $\Pi=0$ and we recover the metric tensor for the Interior Schwarzschild spacetime.

\begin{figure}
\centering
\includegraphics[width=1\columnwidth]{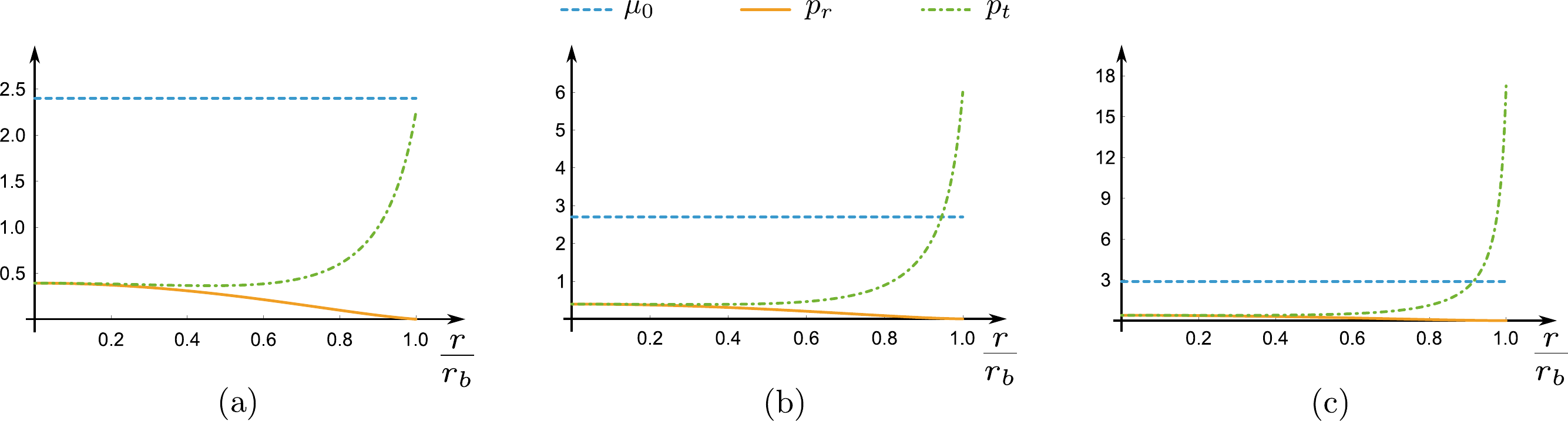}

\caption{\label{Figure:Generalized_In_Schw_plots}Energy density, radial and
tangential pressures as functions of the radial coordinate, for various
values of the compactness parameter for the solution~\ref{Compactness_eq:Generalized_Interior_Schwarzschild_metric}.
In all cases, we have set $A=1$ and $B=0.5$. In (a) $M/r_b=0.4$,
(b) $M/r_b=0.45$ and (c) $M/r_b=0.48$.}
\end{figure}

A spacetime
characterized by~\eqref{Compactness_eq:Generalized_Interior_Schwarzschild_metric}
is nonphysical, since it would describe a star with a homogeneous
energy density. However, it has a number of interesting
properties. For instance, it is always possible to choose values of  $B<8/9$ such that regular solutions with compactness arbitrarily close to
the black hole limit: $r_b=2M$, exist.
Additionally, for any value of the
compactness up to the black hole limit, it is always possible to find
values of $B$ such that all matter variables are non-negative and,
trivially, the pointwise weak and strong energy conditions are verified. To exemplify this property, in 
Figure~\ref{Figure:Generalized_In_Schw_plots}, we present the
behavior of the matter variables for various values of the compactness
parameter and a fixed value of $B$. Notice that the tangential pressure does
not have an asymptote at $r=r_b$, unless $r_b=2M$.

A thorough discussion of the solution~\eqref{Compactness_eq:Generalized_Interior_Schwarzschild_metric}
has been presented elsewhere \cite{Luz:2026fzu}. For the purpose of this article, we will make
use of this solution to study the maximum compactness of a stellar
compact object composed of a non-perfect fluid. We have analyzed the
radial stability of the new solution using the new unified perturbation framework, considering the anisotropic ansatze following from the Eckart theory,
the BDNK fluid model, and the Truncated Israel-Stewart theory, with
shear viscosity and relaxational time modeled by Eq.~\eqref{Eckart_eq:Shear_viscosity_ansatz}.
For fixed values of the compactness parameter, we have found that
the maximum absolute value of the fundamental eigenfrequency occurs
for $B=0$, that is, when the radial pressure is zero. This, in itself,
is an interesting result and in line with previous reports asserting
that a higher value of the tangential pressure, relative to the radial
pressure, yields a more stable configuration~\citep{Mondal_Bagchi_2024}.
In the absence of radial pressure, the unperturbed stellar compact
object is completely held against collapse by tangential stresses.
Considering Eq.~\eqref{Eckart_eq:Shear_viscosity_ansatz}, we see
that, for $B=0$, shear viscosity, $\eta$, and the relaxational time,
$\tau_{2}$, both vanish. This is the expected result. Shear viscosity
is a consequence of the internal friction between fluid layers moving
at different velocities. For vanishing radial pressure, the fluid
layers have no direct thermodynamic interaction, hence the viscosity
and the relaxational time should be zero. In that case, Eq.~\eqref{Comoving_Perturbation_eqs:Final_system_P_constraint}
implies that
$\Psi_{\mathcal{P}}^{\left(\upsilon\right)}=0$, hence $\dot{\Pi}=0$. That is, the anisotropic pressure of the perturbed fluid is unchanged and
is given simply by its unperturbed value.

Since the unperturbed spacetime is characterized by a constant
energy density, assuming the perturbed fluid verifies a barotropic
equation of state: $p+f\left(\mu\right)=0$, we have $f_{p}=1$ and
$f_{\mu}=-c_{s}^{2}$, where, at linear level, $c_{s}^{2}$ is the,
constant, adiabatic speed of sound. Imposing causality, that is, $c_{s}^{2}\leq1$,
we can find that the maximum value of the compactness parameter, beyond
which the family of solutions~\eqref{Compactness_eq:Generalized_Interior_Schwarzschild_metric}
is dynamically unstable. Computing the fundamental eigenfrequency
for various values of the compactness parameter, we have found that

\begin{equation}
\begin{aligned}
	 \frac{M}{r_{b}}=0.4192 & \Rightarrow\lambda_{0}^{2}>0\,,
	& \frac{M}{r_{b}}=0.4193 & \Rightarrow\lambda_{0}^{2}<0\,,
	\end{aligned}
\end{equation}
that is, for compactness higher than $0.4193$, the fundamental eigenfrequency,
$\lambda_{0}$, of any solution of the form~\eqref{Compactness_eq:Generalized_Interior_Schwarzschild_metric}
is imaginary, hence the spacetime is dynamically unstable. As we have
stated above, for $B=0$, Eq.~\eqref{Compactness_eq:Generalized_Interior_Schwarzschild_metric}
characterizes a spherically symmetric, non-trivial solution
of the Einstein field equations with constant energy density, identically zero radial pressure
and zero cosmological constant. Assuming that this solution represents
the extreme scenario for a static, spherically symmetric, self-gravitating
non-dissipative fluid, we can use it to conjecture the maximum compactness of a star, beyond which
it becomes dynamically unstable. With these assumptions, the perturbation equations derived above allow us to conclude that a static, spherically symmetric solution of the Einstein field
equations for a self-gravitating, non-dissipative matter fluid is
linearly dynamically unstable if
\begin{equation}
\frac{M}{r_{b}}\apprge0.4193\,.\label{Compactness_eq:Upper_bound}
\end{equation}

Remarkably, the limit above is in agreement with the results of Ref.~\citep{Raposo_et_al_2019},
found when studying linear perturbations of a particular type of anisotropic
solutions of GR and considering extremely high anisotropic configurations.
The fact that these results are in such good agreement shows that our initial assumption is indeed solid. 

We stress, however, that the upper bound~\eqref{Compactness_eq:Upper_bound}  depends critically on the model of anisotropy employed. The previous result was derived considering the anisotropic ansatze following from
the Eckart theory, the BDNK fluid model, and the Truncated Israel-Stewart
theory. It is, however, possible to consider a different ansatz or other models for the shear viscosity,
$\eta$, or the relaxational time, $\tau_{2}$, than the ones in Eq.~\eqref{Eckart_eq:Shear_viscosity_ansatz},
such that these quantities do not vanish for zero radial pressure.
Such modifications are expected to yield a distinct upper bound on compactness.

\section{Conclusion}\label{Sec:Conclusion}

We have developed a covariant, gauge-invariant set of linear equations
for radial adiabatic perturbations of stellar compact objects. The
equation of state and the anisotropic ansatz were codified into two
generic functions and integrated into the perturbation equations, such
that the final system is applicable to very generic setups, without
the need to rederive the perturbation equations depending on the
thermodynamic model.

The new set of equations was explicitly written for the most studied anisotropic ansatze,  namely, the Eckart theory, the BDNK
fluid model, and the Truncated Israel-Stewart theory. We have shown
that in the context of radial adiabatic perturbations, the Eckart
and the BDNK theories are essentially equivalent at the level of linear perturbations,
differing only in the transport coefficients that characterize the fluid. We have found the
general form of the solutions around the center for all possible forms
of the equation of state.

As a first application of the equations, we have considered the perturbation of some classical spacetime solutions, studying the effects
of shear viscosity and the relaxational time on the perturbed eigenmodes. We have found that
for higher values of shear viscosity, there is a significant difference
of the behavior of the eigenfunctions and eigenfrequencies when considering
the Eckart and the BDNK fluid models, compared to considering the
Truncated Israel-Stewart. 
The results somewhat reduce our confidence in the use of
 the Eckart and the BDNK theories
for fluids out of thermodynamic equilibrium to model radial, adiabatic
perturbations of stellar compact objects. The key issue is that these theories do not predict
a relaxation mechanism for anisotropic stresses. Without such a mechanism, for higher values of shear viscosity, the amplitude
of the perturbations around the center can be many orders of magnitude
smaller than the amplitude of the perturbations at the boundary of
the star. This effect is present for all eigenmodes but becomes increasingly
more prominent for higher-order eigenmodes. Nonetheless, we remark
that this result, by itself, does not mean that the Eckart or the
BDNK models should be disregarded. Here, we have considered only adiabatic perturbations,
that is, we have imposed that there is no dissipation within the fluid.
This hypothesis is, of course, an approximation and imposes very strong
constraints on the fluid models. Allowing dissipative flows may prevent the effect of amplification of the amplitude of the perturbations
toward the outer boundary predicted by the partial Eckart theory or
the partial BDNK theory.

Next, we have considered the analysis of the stability of relativistic stellar objects in which 
the source is described by the MIT bag model: the so-called Strange stars. The formalism we have constructed allowed us to analyze for the first time the perturbations of this type of object by including the anisotropic 
ansatze suggested by the non-equilibrium thermodynamic theories. The results on the eigenfrequencies we obtain are consistent with those found in the literature, even imposing other ansatze. However, we also found an indication that strange stars might be inherently non-perturbative, as their perturbations appear to have very high amplitudes. The lesson here is that, in considering more realistic models of relativistic stars, one also needs to care about the properties of the eigenfunctions, not just the eigenvalues, as the behavior of the former can contain relevant physical information.

Lastly, we have used the new perturbation scheme to conjecture on
the maximum value of the compactness parameter beyond which any stellar compact object will be linearly, dynamically unstable. In that regard,
we employed a solution of the Einstein field equations
with constant energy density that is regular for compactness values
up to the black hole limit, and we studied its dynamical stability under
radial adiabatic perturbations, considering the Eckart, BDNK, and Truncated
Israel-Stewart theories. We have found that, for this solution, maximum
stability occurs when radial pressure vanishes identically. This result
is important since it implies that the value found for the maximum
compactness is independent of the value of shear viscosity and the
relaxational time, and consequently, also independent of the anisotropic
ansatz. Assuming the perturbed fluid is characterized by a barotropic
equation of state, we have found the upper bound $M/r_{b}\approx0.4193$.
This value is in very good agreement with values reported in the literature
when considering specific fluid models with extremely high anisotropies.

\begin{acknowledgments}
PL thanks the Funda\c{c}\~{a}o para a Ci\^{e}ncia e Tecnologia (FCT), Portugal, for the financial support to the Center for Astrophysics and Gravitation through grant No. UID/PRR/00099/2025 and grant No. UID/00099/2025. The research activities of SC have been carried out in the framework of the INFN Research Project QGSKY.
\end{acknowledgments}

\appendix

\section{Equations of state for the anisotropies\label{Appendix_sec:EoS2}}

As discussed in Subsection~\ref{subsec:Comoving_complete_matter_model},
in addition to the equation of state, relating the matter fluid variables,
$f\left(\mu,p,\Pi\right)=0$, another equation has to be provided
that characterizes the source of the anisotropic pressure. Quite generally,
we have considered an equation of the form~\eqref{Comoving_Perturbation_eqs:EoS_2_differential_general_full}.
Here we will show explicitly how various popular models in the literature
can be cast in the form of Eq.~\eqref{Comoving_Perturbation_eqs:EoS_2_differential_general_full}
and the gauge invariant form~\eqref{Comoving_Perturbation_eqs:EoS_2_differential_general_linear_derivatives}.

To write the various ansatz introduced in the literature in the language
of the covariant 1+1+2 formalism, let

\begin{equation}
\left\{ \begin{aligned}p_{r} & =p+\Pi\\
p_{t} & =p-\frac{1}{2}\Pi
\end{aligned}
\right.\Leftrightarrow\left\{ \begin{aligned}p_{r}-p_{t} & =\frac{3}{2}\Pi\\
2p_{t}+p_{r} & =3p
\end{aligned}
\right.\,,\label{Appendix_ansatze_eq:Radial_tangential_pressures_p_Pi_relations}
\end{equation}
define the radial and tangential components of a fluid's pressure,
respectively, $p_{r}$ and $p_{t}$.

\subsection{Quasi-local ansatz}

The so-called quasi-local ansatz introduced by Horvat et al.~\citep{Horvat_et_al_2011}
reads
\begin{equation}
p_{t}-p_{r}=\zeta_{H}\beta p_{r}\Leftrightarrow\frac{3}{2}\Pi=-\zeta_{H}\beta\left(p+\Pi\right)\,,\label{Appendix_ansatze_eq:QuasiLocal_EoS}
\end{equation}
where $\zeta_{H}\in\mathbb{R}$ and $\beta=\beta\left(x^{\alpha}\right)$
is a function of the spacetime representing the compactness of the
star at a given circumferential radius. In a static, spherically symmetric spacetime
the compactness can be written covariantly in terms of the 1+1+2 variables:
\begin{equation}
\beta=\frac{\mathcal{A}\phi-p+\Lambda-\Pi}{\frac{1}{4}\phi^{2}+\mathcal{A}\phi-p+\Lambda-\Pi}\,.
\end{equation}

Taking the dot derivative, wecan obtian an equivalent relation which is gauge invariant:
\begin{equation}
\begin{aligned}\left[\frac{\zeta_{H}\left(p+\Pi\right)\phi^{2}}{4\left(\frac{1}{4}\phi^{2}+\mathcal{A}\phi-p+\Lambda-\Pi\right)^{2}}-\zeta_{H}\beta\right]\mathsf{p}+\left[\frac{\zeta_{H}\left(p+\Pi\right)\phi^{2}}{4\left(\frac{1}{4}\phi^{2}+\mathcal{A}\phi-p+\Lambda-\Pi\right)^{2}}-\zeta_{H}\beta-\frac{3}{2}\right]\mathcal{P}\\
-\frac{\zeta_{H}\left(p+\Pi\right)\phi^{3}}{4\left(\frac{1}{4}\phi^{2}+\mathcal{A}\phi-p+\Lambda-\Pi\right)^{2}}\mathsf{A}+\frac{\zeta_{H}\left(p+\Pi\right)\left(\phi\mathcal{A}-2p+2\Lambda-2\Pi\right)\phi}{4\left(\frac{1}{4}\phi^{2}+\mathcal{A}\phi-p+\Lambda-\Pi\right)^{2}}\mathsf{F} & =0\,.
\end{aligned}
\end{equation}
Therefore, the quasi-local ansatz corresponds to a particular case
of Eq.~\eqref{Comoving_Perturbation_eqs:EoS_2_algebric_general_linear_derivatives}
with the following non-trivial coefficients:
\begin{equation}
\begin{aligned}g_{p} & =\frac{\zeta_{H}\left(p+\Pi\right)\phi^{2}}{4\left(\frac{1}{4}\phi^{2}+\mathcal{A}\phi-p+\Lambda-\Pi\right)^{2}}-\zeta_{H}\beta\,, & g_{\Pi} & =\frac{\zeta_{H}\left(p+\Pi\right)\phi^{2}}{4\left(\frac{1}{4}\phi^{2}+\mathcal{A}\phi-p+\Lambda-\Pi\right)^{2}}-\zeta_{H}\beta-\frac{3}{2}\,,\\
g_{\mathcal{A}} & =-\frac{\zeta_{H}\left(p+\Pi\right)\phi^{3}}{4\left(\frac{1}{4}\phi^{2}+\mathcal{A}\phi-p+\Lambda-\Pi\right)^{2}}\,, & g_{\phi} & =\frac{\zeta_{H}\left(p+\Pi\right)\left(\phi\mathcal{A}-2p+2\Lambda-2\Pi\right)\phi}{4\left(\frac{1}{4}\phi^{2}+\mathcal{A}\phi-p+\Lambda-\Pi\right)^{2}}\,.
\end{aligned}
\end{equation}

\subsection{Bowers-Liang ansatz}

The model derived by Bowers and Liang for a constant-density anisotropic
star~\citep{Bowers_Liang_1974} follows from considering the non-linear
relation between the matter variables:
\begin{equation}
p_{t}-p_{r}=\frac{4}{\phi^{2}}\zeta_{BL}\left(\mu+p_{r}\right)\left(\mu+3p_{r}\right)\Leftrightarrow\Pi=-\frac{8}{3\phi^{2}}\zeta_{BL}\left(\mu+p+\Pi\right)\left(\mu+3p+3\Pi\right)\,,
\end{equation}
where $\zeta_{BL}\in \mathbb{R}$. Taking the dot-derivative, we find
\begin{equation}
\begin{aligned}\frac{16\zeta_{BL}}{3\phi^{3}}\left(\mu+p+\Pi\right)\left(\mu+3p+3\Pi\right)\mathsf{F}-\frac{16\zeta_{BL}}{3\phi^{2}}\left(\mu+2p+2\Pi\right)\mathsf{m}\\
-\frac{16\zeta_{BL}}{3\phi^{2}}\left(2\mu+3p+3\Pi\right)\mathsf{p}-\left[\frac{16\zeta_{BL}}{3\phi^{2}}\left(2\mu+3p+3\Pi\right)+1\right]\mathcal{P} & =0\,.
\end{aligned}
\end{equation}
Hence, the Bowers-Liang ansatz corresponds to a particular case of
Eq.~\eqref{Comoving_Perturbation_eqs:EoS_2_algebric_general_linear_derivatives}
with the following non-trivial coefficients:
\begin{equation}
\begin{aligned}g_{\mu} & =-\frac{16\zeta_{BL}}{3\phi^{2}}\left(\mu+2p+2\Pi\right)\,, & g_{p} & =-\frac{16\zeta_{BL}}{3\phi^{2}}\left(2\mu+3p+3\Pi\right)\,,\\
g_{\Pi} & =-\left[\frac{16\zeta_{BL}}{3\phi^{2}}\left(2\mu+3p+3\Pi\right)+1\right]\,, & g_{\phi} & =\frac{16\zeta_{BL}}{3\phi^{3}}\left(\mu+p+\Pi\right)\left(\mu+3p+3\Pi\right)\,.
\end{aligned}
\end{equation}

Together with the field equations, the Bowers-Liang ansatz imposes strong constraints on the anisotropic
pressure: $\Pi$ must vanish and have a critical point at the center.

\subsection{Herrera-Barreto ansatz}

The model formalized by Herrera and Barreto~\citep{Herrera_Barreto_2013}
reads
\begin{equation}
p_{t}-p_{r}= \frac{\left(\zeta_{HB}-1\right)r}{2\zeta_{HB}} \frac{dp_{r}}{dr} \Leftrightarrow \Pi=-\frac{2\left(\zeta_{HB}-1\right)}{3\zeta_{HB}\phi} \left(\widehat{p}+\widehat{\Pi}\right)\,,
\label{Appendix_ansatze_eq:Herrera_Barreto_EoS}
\end{equation}
where $\zeta_{HB}\in\mathbb{R}$.

Using the conservation law, in the absence of dissipative heat flows
and setting vector and tensor components to zero,
\begin{equation}
\widehat{p}+\widehat{\Pi}=-\frac{3}{2}\phi\Pi-\left(\mu+p+\Pi\right)\mathcal{A}\,,\label{Appendix_ansatze_eq:Conservation_law}
\end{equation}
equation~\eqref{Appendix_ansatze_eq:Herrera_Barreto_EoS} can be
written as the algebraic equation
\begin{equation}
\left[3\phi+2\left(1-\zeta_{HB}\right)\mathcal{A}\right]\Pi=2\left(\zeta_{HB}-1\right)\left(\mu+p\right)\mathcal{A}\,.
\end{equation}
Taking the dot-derivative
yields
\begin{equation}
\begin{aligned}\left[3\phi+2\left(1-\zeta_{HB}\right)\mathcal{A}\right]\mathcal{P}+2\left(1-\zeta_{HB}\right)\mathcal{A}\mathsf{m}+2\left(1-\zeta_{HB}\right)\mathcal{A}\mathsf{p}\\
+3\Pi\mathsf{F}+2\left(1-\zeta_{HB}\right)\left(\mu+p+\Pi\right)\mathsf{A} & =0\,.
\end{aligned}
\end{equation}
Therefore, the Herrera-Barreto ansatz corresponds to a particular
case of Eq.~\eqref{Comoving_Perturbation_eqs:EoS_2_algebric_general_linear_derivatives}
with the following non-trivial coefficients:
\begin{equation}
\begin{aligned}g_{\mu} & =2\left(1-\zeta_{HB}\right)\mathcal{A}\,, & g_{p} & =2\left(1-\zeta_{HB}\right)\mathcal{A}\,, & g_{\phi} & =3\Pi\,,\\
g_{\Pi} & =3\phi+2\left(1-\zeta_{HB}\right)\mathcal{A}\,, & g_{\mathcal{A}} & =2\left(1-\zeta_{HB}\right)\left(\mu+p+\Pi\right)\,.
\end{aligned}
\end{equation}

The Herrera-Barreto anisotropic ansatz, Eq.~\eqref{Appendix_ansatze_eq:Herrera_Barreto_EoS},
together with the field equations implies that at the center $\partial_{r}p_{t}=\partial_{r}p_{r}=0$. Therefore, like the Bowers-Liang ansatz, it constrains $\Pi$ to vanish and have a critical point at the center.

\subsection{Covariant ansatz}

The anisotropic ansatz introduced by Raposo et al. in Ref.~\citep{Raposo_et_al_2019},
has been dubbed in the literature as the covariant ansatz, due to
its manifestly covariant form. The ansatz reads
\begin{equation}
p_{r}-p_{t}=\zeta_R h\left(\mu\right)\widehat{p}_r \Leftrightarrow\frac{3}{2}\Pi=\zeta_R h\left(\mu\right)\left(\widehat{p}+\widehat{\Pi}\right)\,, \label{Appendix_ansatze_eq:Raposo_et_al_EoS}
\end{equation}
where $\zeta_R\in\mathbb{R}$ and $h\left(\mu\right)$ is an arbitrary
differentiable function of the energy density, $\mu$. Using the conservation
law~\eqref{Appendix_ansatze_eq:Conservation_law}, Eq.~\eqref{Appendix_ansatze_eq:Raposo_et_al_EoS}
can be written in algebraic form:
\begin{equation}
\frac{3}{2}\left[\zeta_{R}h\left(\mu\right)\phi+1\right]\Pi+\zeta_{R}h\left(\mu\right)\left(\mu+p+\Pi\right)\mathcal{A}=0\,.\label{Appendix_ansatze_eq:Raposo_et_al_algebraic}
\end{equation}
Taking the dot-derivative, we find
\begin{equation}
\begin{aligned}\left[\zeta_{R}h\left(\mu\right)\left(\phi+\frac{2}{3}\mathcal{A}\right)+1\right]\mathcal{P}+\frac{2}{3}\zeta_{R}\left[\frac{3}{2}\Pi\phi h_{\mu}+\mathcal{A}\left(\mu+p+\Pi\right)h_{\mu}+\mathcal{A}h\left(\mu\right)\right]\mathsf{m}\\
+\frac{2}{3}\zeta_{R}\mathcal{A}h\left(\mu\right)\mathsf{p}+\frac{2}{3}\zeta_{R}h\left(\mu\right)\left(\mu+p+\Pi\right)\mathsf{A}+\zeta_{R}\Pi h\left(\mu\right)\mathsf{F} & =0\,,
\end{aligned}
\end{equation}
where $h_{\mu}\equiv\partial_{\mu}h$. Hence, the covariant ansatz
corresponds to a particular case of Eq.~\eqref{Comoving_Perturbation_eqs:EoS_2_algebric_general_linear_derivatives}
with the following non-trivial coefficients:
\begin{equation}
\begin{aligned}g_{\mu} & =\frac{2}{3}\zeta_{R}\left[\frac{3}{2}\Pi\phi h_{\mu}+\mathcal{A}\left(\mu+p+\Pi\right)h_{\mu}+\mathcal{A}h\left(\mu\right)\right]\,, & g_{p} & =\frac{2}{3}\zeta_{R}\mathcal{A}h\left(\mu\right)\,, & g_{\phi} & =\zeta_{R}\Pi h\left(\mu\right)\,,\\
g_{\Pi} & =\zeta_{R}h\left(\mu\right)\left(\phi+\frac{2}{3}\mathcal{A}\right)+1\,, & g_{\mathcal{A}} & =\frac{2}{3}\zeta_{R}h\left(\mu\right)\left(\mu+p+\Pi\right)\,.
\end{aligned}
\end{equation}

Like the Bowers-Liang and the Herrera-Barreto ansatze, the covariant ansatz imposes strong constraints on the fluid model.
The field equations and relation~\eqref{Appendix_ansatze_eq:Raposo_et_al_EoS} imply that, at the center of the star, either $\zeta_{R}h\left(\mu\right)$
vanishes, or the anisotropic pressure,
$\Pi$, is zero and must have a critical point.

\subsection{Eckart theory and the BDNK effective theory}

The Eckart theory for non-equilibrium thermodynamics provides the
following relation between the anisotropic pressure scalar and the
shear scalar:
\begin{equation}
\Pi=-2\eta\Sigma\,,\label{Appendix_ansatze_eq:Eckart_EoS}
\end{equation}
where $\eta$ represents the shear viscosity of the fluid and is assumed
to be a function of the matter variables. Taking the dot-derivative
of this equation yields
\begin{equation}
\mathcal{P}+2\dot{\eta}\Sigma+2\eta\dot{\Sigma}=0\,.
\end{equation}

In the context of perturbation theory, the term $2\dot{\eta}\Sigma$
is higher than linear order with respect to a static background. Then,
at linear level, the Eckart theory is characterized by the relation
\begin{equation}
\mathcal{P}+2\eta\dot{\Sigma}=0\,.
\end{equation}
This relation corresponds to a particular case of Eq.~\eqref{Comoving_Perturbation_eqs:EoS_2_algebric_general_linear_derivatives}
with the following non-trivial coefficients:
\begin{equation}
\begin{aligned}g_{\Pi} & =1\,, & g_{\Sigma} & =2\eta\,.\end{aligned}
\end{equation}

Recently, a causal relativistic first-order fluid model was proposed
by Bemfica, Disconzi, Noronha, and Kovtun~\citep{Bemfica_Disconzi_Noronha_2018,Bemfica_Disconzi_Noronha_2019,Kovtun_2019,Bemfica_Disconzi_Noronha_2022},
aiming to solve many of the issues of causal relativistic non-equilibrium
thermodynamic theories. The full theory is quite different from Eckart's
theory, however, in LRS spacetimes, both theories imply Eq.~\eqref{Appendix_ansatze_eq:Eckart_EoS}.
Therefore, the Eckart theory and the BDNK theory can be considered
to yield the same anisotropic ansatz and the dynamical behavior of
linearized, radial adiabatic perturbations of fluids, verifying the
structure equation of either theory will be the same. Nonetheless,
since the remaining structure equations of the Eckart theory and BDNK
theory are not equal, the thermodynamic behavior of the fluid is of
course different. In particular, in the Eckart theory, $\mathsf{p}$
is related to the derivative of the expansion scalar and the bulk
viscosity, whereas, in the BDNK theory, $\mathsf{p}$ is related with
various kinematical and matter variables through the bulk viscosity
and a transport parameter.

\subsection{Truncated Israel-Stewart theory for non-equilibrium thermodynamics}

The Truncated Israel-Stewart theory for non-equilibrium thermodynamics,
also called the Maxwell-Cattaneo form of Israel-Stewart theory, \citep{Israel_1976,Israel_Stewart_1979}
yields the following differential equation for the anisotropic pressure
\begin{equation}
\tau_{2}h^{\mu}{}_{\alpha}h^{\nu}{}_{\beta}\dot{\pi}_{\mu\nu}+\pi_{\alpha\beta}=-2\eta\sigma_{\alpha\beta}\,,
\end{equation}
where $\tau_{2}$ is a relaxational time associated with the anisotropic
stresses and $h^{\mu}{}_{\alpha}=g^{\mu}{}_{\alpha}+u^{\mu}u_{\alpha}$.
Separating the scalar components of $\pi_{\alpha\beta}$, this equation
implies
\begin{equation}
\tau_{2}\dot{\Pi}+\Pi=-2\eta\Sigma\,.
\end{equation}
Taking the dot-derivative and assuming that the relaxational time,
$\tau_{2}$, and the shear viscosity, $\eta$, are functions of the
matter variables, in the context of perturbation theory, we find
\begin{equation}
\tau_{2}\ddot{\Pi}+\dot{\Pi}=-2\eta\dot{\Sigma}\,,
\end{equation}
Therefore, the Truncated Israel-Stewart theory for non-equilibrium
thermodynamics corresponds to a particular case of Eq.~\eqref{Comoving_Perturbation_eqs:EoS_2_coeff}
with the following non-trivial coefficients:
\begin{equation}
\begin{aligned}g_{\Pi} & =\left(1+i\upsilon\tau_{2}\right)\,, & g_{\Sigma} & =2\eta\,.\end{aligned}
\end{equation}

\section{Auxiliary quantities for the comoving frame\label{Appendix_sec:Auxiliary_functions_comoving_frame}}

To write the general perturbation equations in the comoving frame
in Section~\ref{subsec:Complete-perturbation-system}, we have introduced
a number of auxiliary functions. Given their length, we present them
separately from the main body of the text:~~~~
\begin{equation}
	\begin{aligned}
		\mathcal{F}_{1}
		& 	=\left[\frac{2}{3}\left(\frac{2\Pi_{0}}{\mu_{0}+p_{0}}-1\right)\frac{f_{\Pi}}{f_{\mu}}\mathcal{A}_{0}+\frac{3}{2}\phi_{0}+\mathcal{A}_{0}+\left(\frac{\left(2\mathcal{E}_{0}+\Pi_{0}\right)f_{\Pi}}{2\left(\mu_{0}+p_{0}\right)f_{\mu}}+\frac{3}{2}\right)\frac{g_{\mathcal{A}}}{g_{\Pi}}\right]\frac{g_{\Sigma}}{g_{\Pi}}i\upsilon\\
		& +\frac{2\Pi_{0}^{2}\mathcal{A}_{0}}{\mu_{0}+p_{0}}-\frac{9}{4}\Pi_{0}\phi_{0}-\left(\mu_{0}+p_{0}-\Pi_{0}\right)\mathcal{A}_{0}+D_{e}\left(\frac{g_{\Sigma}}{g_{\Pi}}\right)i\upsilon\\
 		& +\frac{3g_{\mathcal{A}}}{2g_{\Pi}}\left(\frac{1}{3}\mu_{0}+2p_{0}-\frac{5}{3}\Lambda-2\mathcal{A}_{0}\phi_{0}-\mathcal{A}_{0}^{2}+\Pi_{0}-\upsilon^{2}\right)\\
		& -\frac{g_{\mathcal{A}}\left(\mu_{0}-3p_{0}+4\Lambda+6\mathcal{A}_{0}\phi_{0}-6\Pi_{0}\right)\Pi_{0}}{4g_{\Pi}\left(\mu_{0}+p_{0}\right)}\,,
	\end{aligned}
\end{equation}
\begin{equation}
\begin{aligned}\mathcal{F}_{2} & =\frac{g_{\mathcal{A}}}{g_{\Pi}}\left[\left(\frac{\left(2\mathcal{E}_{0}+\Pi_{0}\right)f_{\Pi}}{2\left(\mu_{0}+p_{0}\right)f_{\mu}}+\frac{3}{2}\right)\frac{g_{p}}{g_{\Pi}}-\frac{\left(2\mathcal{E}_{0}+\Pi_{0}\right)f_{p}}{2\left(\mu_{0}+p_{0}\right)f_{\mu}}\right]-2\mathcal{A}_{0}-\frac{2}{3}\left(\frac{2\Pi_{0}}{\mu_{0}+p_{0}}-1\right)\frac{f_{p}}{f_{\mu}}\mathcal{A}_{0}\\
 & +\left[\frac{2}{3}\left(\frac{2\Pi_{0}}{\mu_{0}+p_{0}}-1\right)\frac{f_{\Pi}}{f_{\mu}}\mathcal{A}_{0}+\frac{3}{2}\phi_{0}+2\mathcal{A}_{0}\right]\frac{g_{p}}{g_{\Pi}}+D_{e}\left(\frac{g_{p}}{g_{\Pi}}\right)\,,
\end{aligned}
\end{equation}

\begin{equation}
\begin{aligned}\mathcal{F}_{3} & =\left[\frac{2}{3}\left(\frac{2\Pi_{0}}{\mu_{0}+p_{0}}-1\right)\frac{f_{\Pi}}{f_{\mu}}\mathcal{A}_{0}+2\phi_{0}-\mathcal{A}_{0}\right]\frac{g_{\mathcal{A}}}{g_{\Pi}}+D_{e}\left(\frac{g_{\mathcal{A}}}{g_{\Pi}}\right)\\
 & +\left(\frac{\left(2\mathcal{E}_{0}+\Pi_{0}\right)f_{\Pi}}{2\left(\mu_{0}+p_{0}\right)f_{\mu}}+\frac{3}{2}\right)\left(\frac{g_{\mathcal{A}}}{g_{\Pi}}\right)^{2}-\mu_{0}-p_{0}-\Pi_{0}\,;
\end{aligned}
\end{equation}
\begin{equation}
	\begin{aligned}
		G_{1} & 	=\frac{2}{3}\left[\frac{\left(\widehat{\mu}_{0}+\widehat{p}_{0}\right)f_{\Pi}g_{\Sigma}}{\left(\mu_{0}+p_{0}\right)f_{\mu}g_{\Pi}}+\mathcal{A}_{0}\frac{f_{\Pi}g_{\Sigma}}{f_{\mu}g_{\Pi}}-D_{e}\left(\frac{f_{\Pi}g_{\Sigma}}{f_{\mu}g_{\Pi}}\right)\right]i\upsilon-\frac{f_{\Pi}g_{\mathcal{A}}}{f_{\mu}g_{\Pi}}\left[\frac{\left(2\mathcal{E}_{0}+\Pi_{0}\right)f_{\Pi}}{3\left(\mu_{0}+p_{0}\right)f_{\mu}}+1\right]\frac{g_{\Sigma}}{g_{\Pi}}i\upsilon\\
	 	& +\frac{f_{\Pi}g_{\mathcal{A}}\left(\mu_{0}-3p_{0}+4\Lambda+6\mathcal{A}_{0}\phi_{0}-6\Pi_{0}\right)\Pi_{0}}{6f_{\mu}g_{\Pi}\left(\mu_{0}+p_{0}\right)}-\frac{3}{2}\left(\mu_{0}+p_{0}\right)\phi_{0}-\widehat{\Pi}_{0}+\frac{\Pi_{0}\left(\widehat{\mu}_{0}+\widehat{p}_{0}\right)}{\mu_{0}+p_{0}}\\
	 	& -\frac{f_{\Pi}g_{\mathcal{A}}}{f_{\mu}g_{\Pi}}\left(\frac{1}{3}\mu_{0}+2p_{0}-\frac{5}{3}\Lambda-2\mathcal{A}_{0}\phi_{0}-\mathcal{A}_{0}^{2}+\Pi_{0}-\upsilon^{2}\right)\,,
	\end{aligned}
\end{equation}
\begin{equation}
G_{2}=\frac{2}{3}\left[D_{e}\left(\frac{f_{p}}{f_{\mu}}-\frac{f_{\Pi}g_{p}}{g_{\Pi}f_{\mu}}\right)-\frac{\widehat{\mu}_{0}+\widehat{p}_{0}}{\mu_{0}+p_{0}}\left(\frac{f_{p}}{f_{\mu}}-\frac{f_{\Pi}g_{p}}{f_{\mu}g_{\Pi}}\right)\right]-\frac{f_{\Pi}g_{\mathcal{A}}}{f_{\mu}g_{\Pi}}\left[\frac{2\mathcal{E}_{0}+\Pi_{0}}{3\left(\mu_{0}+p_{0}\right)}\left(\frac{f_{\Pi}g_{p}}{f_{\mu}g_{\Pi}}-\frac{f_{p}}{f_{\mu}}\right)+\frac{g_{p}}{g_{\Pi}}\right]\,,
\end{equation}
\begin{equation}
G_{3}=\frac{2f_{\Pi}g_{\mathcal{A}}}{3f_{\mu}g_{\Pi}}\left[3\mathcal{A}_{0}-\left(\frac{\left(2\mathcal{E}_{0}+\Pi_{0}\right)f_{\Pi}}{2\left(\mu_{0}+p_{0}\right)f_{\mu}}+\frac{3}{2}\right)\frac{g_{\mathcal{A}}}{g_{\Pi}}-\frac{1}{2}\phi_{0}\right]+\frac{2}{3}\left[\frac{\widehat{\mu}_{0}+\widehat{p}_{0}}{\mu_{0}+p_{0}}\left(\frac{f_{\Pi}g_{\mathcal{A}}}{f_{\mu}g_{\Pi}}\right)-D_{e}\left(\frac{f_{\Pi}g_{\mathcal{A}}}{f_{\mu}g_{\Pi}}\right)\right]\,,
\end{equation}
where $D_{e}=\frac{r\phi_{0}}{2}\frac{d}{dr}$, and
\begin{equation}
\mathcal{H}=\mu_{0}+p_{0}+\Pi_{0}+\frac{2g_{\Sigma}}{3f_{\mu}}\left(\frac{f_{\Pi}-f_{p}}{g_{\Pi}-g_{p}}\right)i\upsilon\,.
\label{Appencix_eq:H_function}
\end{equation}
Moreover,
\begin{equation}
	\begin{aligned}\mathcal{M} & =\left[\left(\frac{2\left(\mathcal{U}+\upsilon^{2}\right)f_{\Pi}}{3\left(\mu_{0}+p_{0}\right)f_{\mu}}-1\right)\frac{g_{\mathcal{A}}}{g_{\Pi}}-\phi_{0}\right]^{-1}\left[\frac{2\left(\mathcal{U}+\upsilon^{2}\right)}{3\left(\mu_{0}+p_{0}\right)}\left(\frac{f_{p}}{f_{\mu}}-\frac{f_{\Pi}g_{p}}{f_{\mu}g_{\Pi}}\right)+\frac{g_{p}}{g_{\Pi}}-1\right]\,,\\
		\mathcal{N} & =\left[\left(\frac{2\left(\mathcal{U}+\upsilon^{2}\right)f_{\Pi}}{3\left(\mu_{0}+p_{0}\right)f_{\mu}}-1\right)\frac{g_{\mathcal{A}}}{g_{\Pi}}-\phi_{0}\right]^{-1}\left[\left(\mu_{0}+p_{0}+\Pi_{0}+\frac{2i\upsilon f_{\Pi}g_{\Sigma}}{3f_{\mu}g_{\Pi}}\right)\frac{\mathcal{U}+\upsilon^{2}}{\mu_{0}+p_{0}}-\frac{g_{\Sigma}}{g_{\Pi}}i\upsilon\right]\,.
	\end{aligned}
\end{equation}

Additionally, we provide below the entries of the $\mathds{R}$ and $\Theta$ matrices
introduced in subsec.~\ref{subsec:Breaking_covariance}. Namely,
\begin{equation}
	\begin{aligned}
		\mathds{R}_{11} & =\left[\frac{3g_{p}+4g_{\mathcal{A}}\mathcal{M}}{g_{\Pi}-g_{p}}\left(1+\frac{2g_{\Sigma}}{3\left(g_{\Pi}-g_{p}\right)}\left(\frac{f_{p}}{f_{\mu}}-\frac{f_{\Pi}g_{p}}{f_{\mu}g_{\Pi}}\right)\frac{i\upsilon}{\mathcal{H}}\right)\right]_{r=0}\\
		& -\left[\frac{2i\upsilon g_{\Sigma}}{3\mathcal{H}\left(g_{\Pi}-g_{p}\right)}\frac{f_{\Pi}g_{\mathcal{A}}\mathcal{M}}{f_{\mu}g_{\Pi}}\right]_{r=0}\,,
	\end{aligned}
	\label{Appendix_eq:R11}
\end{equation}
\begin{equation}
	\begin{aligned}
		\mathds{R}_{12} & =\left[\frac{3i\upsilon g_{\Sigma}-4g_{\mathcal{A}}\mathcal{N}}{g_{\Pi}-g_{p}}\left(1+\frac{2g_{\Sigma}}{3\left(g_{\Pi}-g_{p}\right)}\left(\frac{f_{p}}{f_{\mu}}-\frac{f_{\Pi}g_{p}}{f_{\mu}g_{\Pi}}\right)\frac{i\upsilon}{\mathcal{H}}\right)\right]_{r=0}\\
		& +\left[\frac{2i\upsilon g_{\Sigma}}{\mathcal{H}\left(g_{\Pi}-g_{p}\right)}\left(\frac{f_{\Pi}g_{\mathcal{A}}\mathcal{N}}{3f_{\mu}g_{\Pi}}-\frac{3}{2}\left(\mu_{0}+p_{0}\right)\right)\right]_{r=0}\,,
	\end{aligned}
\end{equation}
\begin{equation}
	\begin{aligned}
		\mathds{R}_{21} & =\left[\frac{2g_{\mathcal{A}}\mathcal{M}}{3\mathcal{H}}\left(\frac{4}{g_{\Pi}-g_{p}}\left(\frac{f_{p}}{f_{\mu}}-\frac{f_{\Pi}g_{p}}{f_{\mu}g_{\Pi}}\right)-\frac{f_{\Pi}}{f_{\mu}g_{\Pi}}\right)\right]_{r=0}\\
		& +\left[\frac{2g_{p}}{\mathcal{H}\left(g_{\Pi}-g_{p}\right)}\left(\frac{f_{p}}{f_{\mu}}-\frac{f_{\Pi}g_{p}}{f_{\mu}g_{\Pi}}\right)\right]_{r=0}\,,
	\end{aligned}
\end{equation}
\begin{equation}
	\begin{aligned}
		\mathds{R}_{22} & =\left[\frac{2i\upsilon g_{\Sigma}}{\mathcal{H}\left(g_{\Pi}-g_{p}\right)}\left(\frac{f_{p}}{f_{\mu}}-\frac{f_{\Pi}g_{p}}{f_{\mu}g_{\Pi}}\right)-\frac{3\left(\mu_{0}+p_{0}\right)}{\mathcal{H}}\right]_{r=0}\\
		& -\left[\frac{2g_{\mathcal{A}}\mathcal{N}}{3\mathcal{H}}\left(\frac{4}{g_{\Pi}-g_{p}}\left(\frac{f_{p}}{f_{\mu}}-\frac{f_{\Pi}g_{p}}{f_{\mu}g_{\Pi}}\right)-\frac{f_{\Pi}}{f_{\mu}g_{\Pi}}\right)\right]_{r=0}\,,
	\end{aligned}
	\label{Appendix_eq:R22}
\end{equation}
where we have explicitly stated that all functions that define the $\mathds{R}$ matrix are evaluated at $r=0$, and

\begin{equation}
	\begin{aligned}\Theta_{11} & =\frac{2g_{\Pi}\left(\overline{\mathcal{F}}_{2}+\mathcal{M}\overline{\mathcal{F}}_{3}\right)}{r\phi_{0}\left(g_{\Pi}-g_{p}\right)}\left[1+\frac{2i\upsilon g_{\Sigma}}{3\mathcal{H}\left(g_{\Pi}-g_{p}\right)}\left(\frac{f_{p}}{f_{\mu}}-\frac{f_{\Pi}g_{p}}{f_{\mu}g_{\Pi}}\right)\right]+\frac{2i\upsilon g_{\Sigma}\left(G_{2}+\mathcal{M}\overline{G}_{3}\right)}{r\phi_{0}\mathcal{H}\left(g_{\Pi}-g_{p}\right)}\\
		& +\frac{1}{r}\left\{ \frac{3g_{p}+4g_{\mathcal{A}}\mathcal{M}}{g_{\Pi}-g_{p}}\left[1+\frac{2g_{\Sigma}}{3\left(g_{\Pi}-g_{p}\right)}\left(\frac{f_{p}}{f_{\mu}}-\frac{f_{\Pi}g_{p}}{f_{\mu}g_{\Pi}}\right)\frac{i\upsilon}{\mathcal{H}}\right]\right.\\
		& \qquad\quad \left.-\left[\frac{3g_{p}+4g_{\mathcal{A}}\mathcal{M}}{g_{\Pi}-g_{p}}\left(1+\frac{2g_{\Sigma}}{3\left(g_{\Pi}-g_{p}\right)}\left(\frac{f_{p}}{f_{\mu}}-\frac{f_{\Pi}g_{p}}{f_{\mu}g_{\Pi}}\right)\frac{i\upsilon}{\mathcal{H}}\right)\right]_{r=0}\right\} \\
		& -\frac{1}{r}\left\{ \frac{2i\upsilon g_{\Sigma}}{3\mathcal{H}\left(g_{\Pi}-g_{p}\right)}\frac{f_{\Pi}g_{\mathcal{A}}\mathcal{M}}{f_{\mu}g_{\Pi}}-\left[\frac{2i\upsilon g_{\Sigma}}{3\mathcal{H}\left(g_{\Pi}-g_{p}\right)}\frac{f_{\Pi}g_{\mathcal{A}}\mathcal{M}}{f_{\mu}g_{\Pi}}\right]_{r=0}\right\} \,,
	\end{aligned}
\end{equation}
\begin{equation}
	\begin{aligned}\Theta_{12} & =\frac{2g_{\Pi}\left(\overline{\mathcal{F}}_{1}-\mathcal{N}\overline{\mathcal{F}}_{3}\right)}{r\phi_{0}\left(g_{\Pi}-g_{p}\right)}\left[1+\frac{2i\upsilon g_{\Sigma}}{3\mathcal{H}\left(g_{\Pi}-g_{p}\right)}\left(\frac{f_{p}}{f_{\mu}}-\frac{f_{\Pi}g_{p}}{f_{\mu}g_{\Pi}}\right)\right]+\frac{2i\upsilon g_{\Sigma}\left(\overline{G}_{1}-\mathcal{N}\overline{G}_{3}\right)}{r\phi_{0}\mathcal{H}\left(g_{\Pi}-g_{p}\right)}\\
		& +\frac{1}{r}\left\{ \frac{3i\upsilon g_{\Sigma}-4g_{\mathcal{A}}\mathcal{N}}{g_{\Pi}-g_{p}}\left(1+\frac{2g_{\Sigma}}{3\left(g_{\Pi}-g_{p}\right)}\left(\frac{f_{p}}{f_{\mu}}-\frac{f_{\Pi}g_{p}}{f_{\mu}g_{\Pi}}\right)\frac{i\upsilon}{\mathcal{H}}\right)\right.\\
		& \qquad\quad \left.-\left[\frac{3i\upsilon g_{\Sigma}-4g_{\mathcal{A}}\mathcal{N}}{g_{\Pi}-g_{p}}\left(1+\frac{2g_{\Sigma}}{3\left(g_{\Pi}-g_{p}\right)}\left(\frac{f_{p}}{f_{\mu}}-\frac{f_{\Pi}g_{p}}{f_{\mu}g_{\Pi}}\right)\frac{i\upsilon}{\mathcal{H}}\right)\right]_{r=0}\right\} \\
		& +\frac{1}{r}\left\{ \frac{2i\upsilon g_{\Sigma}}{\mathcal{H}\left(g_{\Pi}-g_{p}\right)}\left(\frac{f_{\Pi}g_{\mathcal{A}}\mathcal{N}}{3f_{\mu}g_{\Pi}}-\frac{3}{2}\left(\mu_{0}+p_{0}\right)\right)\right.\\
		& \qquad\quad \left.-\left[\frac{2i\upsilon g_{\Sigma}}{\mathcal{H}\left(g_{\Pi}-g_{p}\right)}\left(\frac{f_{\Pi}g_{\mathcal{A}}\mathcal{N}}{3f_{\mu}g_{\Pi}}-\frac{3}{2}\left(\mu_{0}+p_{0}\right)\right)\right]_{r=0}\right\} \,,
	\end{aligned}
\end{equation}
\begin{equation}
	\begin{aligned}\Theta_{21} & =\frac{1}{r}\left\{ \frac{2g_{\mathcal{A}}\mathcal{M}}{3\mathcal{H}}\left(\frac{4}{g_{\Pi}-g_{p}}\left(\frac{f_{p}}{f_{\mu}}-\frac{f_{\Pi}g_{p}}{f_{\mu}g_{\Pi}}\right)-\frac{f_{\Pi}}{f_{\mu}g_{\Pi}}\right)\right\} \\
		& \qquad\quad \left.-\left[\frac{2g_{\mathcal{A}}\mathcal{M}}{3\mathcal{H}}\left(\frac{4}{g_{\Pi}-g_{p}}\left(\frac{f_{p}}{f_{\mu}}-\frac{f_{\Pi}g_{p}}{f_{\mu}g_{\Pi}}\right)-\frac{f_{\Pi}}{f_{\mu}g_{\Pi}}\right)\right]_{r=0}\right\} \\
		& +\frac{1}{r}\left\{ \frac{2g_{p}}{\mathcal{H}\left(g_{\Pi}-g_{p}\right)}\left(\frac{f_{p}}{f_{\mu}}-\frac{f_{\Pi}g_{p}}{f_{\mu}g_{\Pi}}\right)-\left[\frac{2g_{p}}{\mathcal{H}\left(g_{\Pi}-g_{p}\right)}\left(\frac{f_{p}}{f_{\mu}}-\frac{f_{\Pi}g_{p}}{f_{\mu}g_{\Pi}}\right)\right]_{r=0}\right\} \\
		& +\frac{2}{r\phi_{0}\mathcal{H}}\left[\frac{2g_{\Pi}}{3\left(g_{\Pi}-g_{p}\right)}\left(\frac{f_{p}}{f_{\mu}}-\frac{f_{\Pi}g_{p}}{f_{\mu}g_{\Pi}}\right)\left(\overline{\mathcal{F}}_{2}+\mathcal{M}\overline{\mathcal{F}}_{3}\right)+G_{2}+\mathcal{M}\overline{G}_{3}\right]\,,
	\end{aligned}
\end{equation}
\begin{equation}
	\begin{aligned}\Theta_{22} & =\frac{2}{r\phi_{0}\mathcal{H}}\left[ \frac{2g_{\Pi}\left(\overline{\mathcal{F}}_{1}-\mathcal{N}\overline{\mathcal{F}}_{3}\right)}{3\left(g_{\Pi}-g_{p}\right)}\left(\frac{f_{p}}{f_{\mu}}-\frac{f_{\Pi}g_{p}}{f_{\mu}g_{\Pi}}\right)+\overline{G}_{1}-\mathcal{N}\overline{G}_{3}\right] \\
		& -\frac{1}{r}\left\{ \frac{3\left(\mu_{0}+p_{0}\right)}{\mathcal{H}}-\left[\frac{3\left(\mu_{0}+p_{0}\right)}{\mathcal{H}}\right]_{r=0}\right\} \\
		& -\frac{1}{r}\left\{ \frac{2g_{\mathcal{A}}\mathcal{N}}{3\mathcal{H}}\left(\frac{4}{g_{\Pi}-g_{p}}\left(\frac{f_{p}}{f_{\mu}}-\frac{f_{\Pi}g_{p}}{f_{\mu}g_{\Pi}}\right)-\frac{f_{\Pi}}{f_{\mu}g_{\Pi}}\right)\right\} \\
		& \qquad\quad \left.-\left[\frac{2g_{\mathcal{A}}\mathcal{N}}{3\mathcal{H}}\left(\frac{4}{g_{\Pi}-g_{p}}\left(\frac{f_{p}}{f_{\mu}}-\frac{f_{\Pi}g_{p}}{f_{\mu}g_{\Pi}}\right)-\frac{f_{\Pi}}{f_{\mu}g_{\Pi}}\right)\right]_{r=0}\right\} \\
		& +\frac{1}{r}\left\{ \frac{2i\upsilon g_{\Sigma}}{\mathcal{H}\left(g_{\Pi}-g_{p}\right)}\left(\frac{f_{p}}{f_{\mu}}-\frac{f_{\Pi}g_{p}}{f_{\mu}g_{\Pi}}\right)-\left[\frac{2i\upsilon g_{\Sigma}}{\mathcal{H}\left(g_{\Pi}-g_{p}\right)}\left(\frac{f_{p}}{f_{\mu}}-\frac{f_{\Pi}g_{p}}{f_{\mu}g_{\Pi}}\right)\right]_{r=0}\right\} \,.
	\end{aligned}
\end{equation}
The barred functions $\overline{\mathcal{F}}_{1,2,3}$
and $\overline{G}_{1,3}$ are defined covariantly in terms of the functions
$\mathcal{F}_{1,2,3}$ and $G_{1,3}$ by the relations:
\begin{equation}
\begin{aligned}\mathcal{F}_{1} & =\frac{3}{2}\frac{g_{\Sigma}}{g_{\Pi}}i\phi_{0}\upsilon+\overline{\mathcal{F}}_{1}\,, & \mathcal{F}_{2} & =\frac{3g_{p}}{2g_{\Pi}}\phi_{0}+\overline{\mathcal{F}}_{2}\,, & \mathcal{F}_{3} & =\frac{2\phi_{0}g_{\mathcal{A}}}{g_{\Pi}}+\overline{\mathcal{F}}_{3}\,,\\
G_{1} & =-\frac{3}{2}\left(\mu_{0}+p_{0}\right)\phi_{0}+\overline{G}_{1}\,, & G_{3} & =-\frac{f_{\Pi}g_{\mathcal{A}}}{3f_{\mu}g_{\Pi}}\phi_{0}+\overline{G}_{3}\,.
\end{aligned}
\end{equation}

\end{document}